\documentclass[aps,prl,twocolumn,preprintnumbers,
superscriptaddress,floatfix]{revtex4}

\usepackage{amssymb,multirow,amsmath}
\usepackage{graphicx,color}

\begin{document}

\newcommand{\Tr}{\mbox{Tr\,}}
\newcommand{\beq}{\begin{equation}}
\newcommand{\eeq}{\end{equation}}
\newcommand{\bea}{\begin{eqnarray}}
\newcommand{\eea}{\end{eqnarray}}
\renewcommand{\Re}{\mbox{Re}\,}
\renewcommand{\Im}{\mbox{Im}\,}

\title{Holographic quark matter with colour superconductivity and a stiff equation of state for compact stars}

\author{Kazem Bitaghsir Fadafan}
\affiliation{ Faculty of Physics, Shahrood University of Technology,
P.O.Box 3619995161 Shahrood, Iran}

\author{Jes\'us Cruz Rojas}
\affiliation{ STAG Research Centre \&  Physics and Astronomy, University of
Southampton, Southampton, SO17 1BJ, UK}

\author{Nick Evans}
\affiliation{ STAG Research Centre \&  Physics and Astronomy, University of
Southampton, Southampton, SO17 1BJ, UK}

\begin{abstract}
We present a holographic model of QCD with a first order chiral restoration phase transition with chemical potential, $\mu$. The first order behaviour follows from allowing a  discontinuity in the dual description as the quarks are integrated out below their constituent mass. The model predicts a deconfined yet massive quark phase at intermediate densities ($350$MeV$<\mu<500$MeV), above the nuclear density phase, which has a very stiff equation of state and a speed of sound close to one.  We also include a holographic description of a colour superconducting condensate in the chirally restored vacuum and study the resulting equation of state. They provides a well behaved first order transition from the   deconfined massive quark phase at very high density ($\mu>500$MeV). We solve the Tolman-Oppenheimer-Volkoff equations with the resulting equations of state and find stable hybrid stars with quark cores. We compute the tidal deformability for these hybrid stars and show they are consistent with LIGO/Virgo data on a neutron star collision. Our holographic model shows that quark matter could be present at the core of such compact stars.
\end{abstract}

\maketitle

\section{I Introduction} \vspace{-0.5cm}

There is a growing literature \cite{Hoyos:2016zke,Ecker:2017fyh,Annala:2017tqz, Jokela:2018ers,Ishii:2019gta,Fadafa:2019euu,Mamani:2020pks,Demircik:2020jkc} attempting to use holography \cite{Maldacena:1997re} to describe the equation of state (EoS) of deconfined quark matter to determine whether it can play a role in neutron star cores. In the first such paper  \cite{Hoyos:2016zke} the exact results at finite density \cite{Kobayashi:2006sb,Karch:2007br}  for the D3/probe D7 system \cite{Karch:2002sh}, which is dual to quark multiplets in a supersymmetric theory, were applied in this context.  This model described a possible deconfined but massive dense quark phase. The resulting EoS was not stiff enough to support quark cores in compact stars such as neutron stars or hybrid stars. 

Recently we adapted this system to a model that included a running anomalous dimension, $\gamma$, for the quark condensate through an effective dilaton profile \cite{Fadafa:2019euu}. That running allowed a description of the chiral restoration transition away from the  deconfined massive phase. The transition occurs dependent upon whether   the Brietenlohner Freedman (BF) bound \cite{Breitenlohner:1982jf} is violated for a scalar in the bulk dual to the chiral condensate. We parameterized the running of $\gamma$ so we could control the derivative of the running at the BF bound violation point. For all choices of the derivative  the model displayed a second order transition from the chirally broken phase to the chirally symmetric phase as $\mu$ increased. Generically the EoS stiffened relative to the pure D3/D7 case with the speed of sound squared in the material  peaking, dependent on the chosen derivative, at $0.6c^2$ for chemical potentials of order the chiral restoration point. Even this was not sufficient to convincingly support large mass neutron stars.

In this paper we present a related model in which the chirally broken phase resists transition to  the chirally restored phase leading to a first order chiral transition.  Prior to the transition the EoS is even stiffer than our previous example with a speed of sound rising close to the speed of light. The key extra ingredient relative to our previous work is that we have allowed a discontinuity in the holographic description at the scale of the IR constituent quark mass, where one might expect the quarks to be integrated out of the running dynamics of the gauge fields. This seems rather natural and it is interesting that our first sensible attempt has led to a very different transition and a much stiffer material. 

Our model remains based on the Dirac Born Infeld (DBI) action of a D7 brane in AdS$_5$ with a scalar field describing the chiral condensate and a U(1) gauge field for the chemical potential. In the spirit of the model in \cite{Alho:2013dka} (motivated by \cite{Jarvinen:2011qe,Alvares:2012kr}) we then include a mass term for the scalar by hand that allows us to include the running  $\gamma$. We input this form from the perturbative QCD running result for $\gamma$ (allowing it to naively extend to the non-perturbative regime). When the scalar mass passes through the BF bound a chiral condensate is induced. This philosophy has been used before in  \cite{Alho:2013dka} to successfully describe $T=\mu=0$ QCD and strongly coupled theories beyond the Standard Model. Previously such models have been able to dodge the question of the deep IR physics below where the quarks become on mass shell as a result of the formation of their constituent mass. Here at finite $\mu$ an explicit description is needed for the IR regime - we propose a simple completion where the anomalous dimension is switched off in this regime and sensible physics results.   

We note that this modelling does not explicitly include confinement - the philosophy is that confinement is a property of the pure Yang Mills theory at scales below the IR constituent quark mass and whilst implied in the model is not directly included.  We assume that as soon as quark density switches on confinement is lost in the plasma. We find a phase with quark density and chiral symmetry breaking - we refer to this phase as a {\it  deconfined massive quark phase}. It is this proposed phase that has a stiff EoS and could play a role in hybrid star cores.

We do not attempt to describe the nuclear physics phase of QCD here. Holography lies close to large $N_c$ where baryons are infinitely massive so it is perhaps not a good starting point (there are good attempts to describe the nuclear phase holographically - for example \cite{Jokela:2018ers,Li:2015uea,Bergman:2007wp,Evans:2012cx} - and one might seek to include those descriptions in the future).  Instead we simply take results from the nuclear physics/neutron star literature \cite{Hebeler:2013nza} that provide three possible varying stiffness EoS for the nuclear phase. Above 308MeV this phase takes over from the vacuum of the holographic theory. In our model though we find that at yet higher density the  deconfined massive quark phase becomes the true vacuum.  Then, the first order transition to the chirally symmetric phase occurs and the stiffness of the EoS falls off sharply. 

The EoS can be inserted into the Tolman-Oppenheimer-Volkoff (TOV) equations to seek pressure versus radius relations inside compact stars (see for example \cite{Haensel_Neutron_Stars1,compactstarbook}). Conditions for stability are discussed in \cite{BardeenThorneMeltzer,Alford:2017vca} which we will review below. Here we show, by solving the TOV equations,  that we do find stable stars with the  deconfined massive quark phase in the core.  This is our first interesting result.

The second task we undertake is to include a holographic description of a possible colour superconducting state \cite{Alford:2007xm} above the first order chiral restoration transition. Traditionally the holography community has declared describing colour superconductivity as very hard because the naive coloured $qq$ order parameter is not gauge invariant and suppressed at large $N_c$ \cite{Shuster:1999tn}. One would need to describe the breaking of the colour group and this remains a tricky issue (see \cite{Faedo:2018fjw} for recent work in this direction). However, in \cite{BitaghsirFadafan:2018iqr}  we proposed to be more cavalier at a phenomenological level and simply allow the inclusion of gauge non-invariant operators and neglect their colour symmetry breaking effects in the dynamics. This was motivated  by the idea that the coloured density of quarks and monopoles (associated with confinement) are already likely to have given Debye masses \cite{Freedman:1976xs} to the gluons before the Cooper pairs form. In this spirit we include a new scalar field dual to the Cooper Pair in analogy to the scalar describing the chiral condensate. Since the Cooper pair carries net baryon charge it couples directly to the U(1) gauge field and the chemical potential itself generates a BF bound violation that can trigger a superconducting phase \cite{Evans:2001ab,Hartnoll:2008vx}.

We construct a bottom up model of the superconducting condensation in the chiral restored phase  - the transition between these two phases is second order at a finite $\mu$ (the value is dependent on the precise coupling used but typical before the true first order transition from the  deconfined massive quark phase). We can describe both a two flavour condensate that breaks the colour symmetry to SU(2) or a three flavour model with a colour flavour locked condensate (although we stress we neglect the impact on the glue sector). The presence of the condensate(s) increases the pressure of the chirally restored phase and pushes the speed of sound squared up. However, for sensible choices of parameters that give a condensate at a scale in the 10s to 100s MeV, $c_s^2\leq 0.5$ and this phase is not stiff enough to serve as the core of a stable hybrid star. Nevertheless the increase in pressure makes the first order transition from the  deconfined massive quark phase to the now superconducting chirally restored phase occur at a lower $\mu$. This serves to complete our model since the speed of sound rises above one in the  deconfined massive phase if allowed to persist to too high $\mu$. 

Thus we present a holographic model that describes a chirally broken vacuum at $\mu=0$. We allow a first order transition to a nuclear phase from 308 MeV, then the holographic model predicts a first order transition to a stiff   deconfined massive quark phase above about 350 MeV before a final first order transition to a superconducting chirally restored phase at around 500 MeV. The EoS still supports the hybrid stars we have previously discussed. This at least  encourages experimental studies of neutron stars to seek such exotic hybrid stars. 

In our final section we also seek to challenge our models of hybrid stars with the recent data from LIGO/Virgo on a neutron star collision \cite{TheLIGOScientific:2017qsa} which has been used to provide constraints on the tidal deformability parameter, $\bar{\lambda}^{(\mathrm{tid})}$, of neutron stars near 1.4 $M_\odot$. We briefly review how to compute $\bar{\lambda}^{(\mathrm{tid})}$ from the TOV equations \cite{Hinderer:2009ca,Postnikov:2010yn,Zhao:2018nyf} and then compute for example equations of state where we have predicted hybrid stars in the appropriate mass range. We find the predictions lie within the allowed region but they could be tested as future data accumulates.

This paper is organised as follows: In section II we review our models of the QCD phases and their EoS - here we review the base D3/D7 model, the nuclear EoS we use, the holographic model of the deconfined massive phase and the chirally restored vacuum, plus finally we add colour superconducting condensates to the chirally restored phase. In section III we then solve the TOV equations to find $M$ versus $R$ relations for hybrid stars. In Section IV we compute the tidal deformability parameter and compare ot the LIGO/Virgo data. In Section V we summarize and conclude.

\vspace{-0.5cm}

\section{II Descriptions of QCD Phases with $\mu$}  \vspace{-0.5cm}

In this section we will work through the descriptions we use of the $\mu=0$ chirally broken vacuum, the nuclear physics phase, a deconfined massive quark phase and a high temperature chirally restored phase with colour superconductivity. All of these descriptions are holographic except for the nuclear phase.  Since our holographic models are inspired by the D3/probe D7 model we review that briefly first. Note that we do not include temperature in any of these discussions since neutron star cores are likely cool. In principle though one could straightforwardly include temperature holographically by allowing a black hole spacetime \cite{Maldacena:1997re}.  \vspace{-0.5cm}

\subsection{IIA  Review of the base D3/D7 probe model}  \vspace{-0.5cm}

At strong 't Hooft coupling $\lambda$ and in the large $N_c$ limit, the dual description of the ${\cal N}=4\,\,SU(N_c)\,\,  SYM$ theory is given by a classical type IIB SUGRA in an $AdS_5 \times S^5$ space time \cite{Maldacena:1997re}. The flavor sector can be introduced as $N_f$ D7 branes extended along the $AdS_5$ geometry and warping an $S^3$ sphere \cite{Karch:2002sh}. At zero temperature the metric background is
\beq
\begin{aligned}
ds^2=&{r^2 \over R^2 }\left(-dt^2+d\vec{x}^2\right)+\\& {R^2 \over r^2} \left( d\rho^2+\rho^2 d\Omega_3^2 +d\chi^2 +\chi^2 d\Omega_1^2\right)
\end{aligned}
\eeq
where $(t,\vec{x})$ are the gauge theory coordinates, the $\rho$ and $\Omega_3$ are on the D7 brane world volume and two transverse directions to the D7 brane are $\chi$ and $\Omega_1$. The energy scale of the boundary theory corresponds to the radial direction $r^2=\rho^2+\chi^2$. The $AdS$ radius is denoted by $R$. To study quarks, consider a D7 probe brane in the background geometry in a quenched approximation when $N_f<<N_c$. There is also a $U(1)$ gauge field $A_a$ where $a=0,1,..,7$ runs over the world volume of the D7 brane. The DBI action for the probe D7 branes is
\begin{equation}
S=-N_f T_{D7}\int d^8 \xi \sqrt{-det \left( g_{ab}+(2\pi\alpha')F_{ab} \right)}\label{S1}
\end{equation}
here $T_{D7}$ is the D7 brane tension,  the world volume coordinates are $\xi_a$, the induced metric is denoted as $g_{ab}$ and $F_{ab}$ is the world volume $U(1)$ gauge field. The two fields we will concentrate on are the gauge field $A_t(\rho)$, with the field strength $F_{\rho t}=A'_t(\rho)$, and the embedding scalar field  $\chi(\rho)$ corresponding to the transverse direction to the D7 brane. They satisfy the following action
at zero temperature
\begin{equation} \label{D7action}
S = - \mathcal{N}_7 \int d \rho\, \rho^3 \sqrt{1 + (\partial_\rho \chi)^2 - (2 \pi \alpha')^2 (\partial_\rho A_t)^2} 
\end{equation}
where $\mathcal{N}_7={N_f } T_{D7} V_3$
and $V_3= 2 \pi^2$ is the volume of the unit $S^3$ on the D7 brane. In the AdS/CFT dictionary $\alpha'^2= {1 \over \lambda}$ where  $\lambda$ is the `t Hooft coupling, $T_{D7} = (2 \pi)^{-7} \alpha^{\prime -4}$ and $\mathcal{N}_7={N_f N_c \lambda \over 16 \pi ^4}$. We divide both sides of the D7 brane action in \eqref{D7action} by the volume of boundary space time $\mathbb{R}^{3,1}$ and henceforth work with action densities.

The holographic interpretation is that the two constants of integration for $\chi$ are the quark mass $m$ and the quark condensate $c$. For $A_t$ we have the chemical potential $\mu$ and the density $d$.

 It is helpful to rescale the factor of $2 \pi \alpha'$ into $A_t$. The constants of integration in the solutions  $\mu$ and $d$ are then on a footing with the constants in $\chi$ ($m$ and $c$) since to move from distances in AdS to energy scales in the field theory one multiplies by $1/2 \pi \alpha'$. Formally one needs to set a scale in the theory by picking for example the IR quark mass' value - after expressing all physical quantities as ratios of this setting scale the $2 \pi \alpha'$ factors then cancel in the ratios. 
  
Thus it is useful to work with the action  (we will reinstate the overall factor of $\mathcal{N}_7$ shortly)
\begin{equation}
L=- \int d \rho\, \rho^3 \sqrt{1 + (\partial_\rho \chi)^2 -(\partial_\rho A_t)^2}
\end{equation}
There are two constants of the motion $d,c$ because only derivatives of $A_t(\rho)$ and $\chi(\rho)$ appear in the Lagrangian. Thus
\begin{equation}
{\partial L \over \partial \chi'}={-\rho^3  \chi'(\rho) \over  \sqrt{1 +  \chi'(\rho)^2 -  A'_t(\rho)^2}}=-c
\end{equation}
and the equation following from  ${\partial L \over  \partial A'_t}$ gives $A_t = d/c ~\chi$.

One can solve for  $A_t'(\rho)$ and $\chi'(\rho)$ in terms of the constants $c$ and $d$ which can be integrated analytically.
The solution is \cite{Karch:2007br}
\begin{equation} \chi =   {c \over 6} (d^2-c^2)^{-1/3}B\left( {\rho^6 \over \rho^6 + d^2 -c^2}; {1 \over 6}; {1 \over 3} \right)
\end{equation}
 with $B$ an incomplete Beta function.
 For $c=d=0$ one finds constant solutions for  $A_t(\rho)$ and $\chi(\rho)$. To obtain the physical solutions for the system with density one needs $d^2-c^2$ positive. Note these solutions have $\chi=A_t=0$ at $\rho=0$ so ``spike'' out of the origin. The density of quarks are D3/D7 strings that pull the D7 to the origin.
 
 The action density evaluated on the solutions is (reinstating $\mathcal{N}_7$)
\begin{equation}
S=-\mathcal{N}_7 \int d\rho \rho^3 \sqrt{\rho^6 \over \rho^6+d^2-c^2}
\end{equation}

One needs a regulator $S_0$ to obtain a finite value for the density action $S$. Numerically, we consider a cutoff at $\Lambda$, a factor of 20 above the IR quark mass, and subtract $S_0=-{\mathcal{N}_7 \Lambda^4 \over 4}$. Then one can define the thermodynamic density free energy by the renormalized action as $\mathcal{F}=-\left(S-S_0\right)$ and as a result an analytic form for the density free energy  \cite{Karch:2007br}
\begin{equation} 
 \label{HoyosF} {\cal F} =- {N_c N_f \over 4 \eta^3 \lambda} (\mu^2 - m^2)^2 , \hspace{0.5cm} \eta = {\Gamma (7/6) \Gamma (1/3) \over \sqrt{\pi}} \end{equation}
To match the asymptotic UV form known from QCD one can pick $\lambda= 3 \pi/\eta^3$ so that:
\begin{equation}  {\cal F} =  {N_c N_f \over 12 \pi^2} \mu^4. \label{uv} \end{equation} 
We will use $N_f=N_c=3$.  

At even infintessimally small temperature, this theory is deconfined. The phase therefore describes a vacuum with a density of quarks of mass $m$. This mass, identified with the constituent quark mass,  must be put in by hand and there is no chiral symmetry breaking mechanism. The EoS, which relates the pressure $P$ to the energy density $\mathcal{E}$ is found from
\begin{equation}
\begin{aligned}
P=-\mathcal{F}, \qquad \mathcal{E}=\mu\frac{\partial P}{\partial \mu}-P.
\end{aligned}
\end{equation}
Here the pressure is too small and the quark interiors of stars can not support neutron stars - see \cite{Hoyos:2016zke}.

\subsection{IIB The $\mu=0$ chiral symmetry breaking phase}

Our base motivation here and in our previous paper \cite{Fadafa:2019euu} is to include the QCD running of the gauge coupling and chiral symmetry breaking into the D3/D7 system to see if the EoS stiffens. In \cite{Fadafa:2019euu}  we added the running as a $\rho$ dependent dilaton prefactor to the action (3). As described in \cite{Alvares:2012kr},  the crucial role the dilaton plays is to provide a running anomalous dimension for the quark bilinear operator which displays as a  $\rho$ dependent mass for the field $\chi$ (after expanding the dilaton). In the previous paper we used a set of functions that ran from $\gamma=0$ in the UV through the critical $\gamma=1$  in the IR (with varying derivative at this point) which indeed triggered chiral symmetry breaking. These models all showed a second order transition from the chirally broken to the chirally symmetric phase. We found the EoS stiffened around the transition so that the speed of sound became as large as 0.6$c$ yet this was still not stiff enough to support hybrid stars.  

Here we will take what appears only a slightly different approach, which is to not introduce running through a dilaton factor but directly through a $\rho$ dependent mass term for $\chi$. This approach has been taken previously in \cite{Alho:2013dka}. Our original motivation for this was that we wanted to add a field for a colour superconducting order parameter in sympathy with $\chi$ but we didn't want that field to experience the same running as $\chi$ which an overall dilaton factor would introduce.  We will see that this ansatz can lead to a yet stiffer EoS. Thus we take the Lagrangian at $\mu=0$
\begin{equation}
    \mathcal{L}=-\rho^3\sqrt{1+(\partial_{\rho}\chi)^2}-\rho  \Delta m^2~ \chi^2
\end{equation}
If we write the additional mass term $\Delta m^{2}$ purely as a function of $\rho$ then were this term to lead to a  violation of the BF bound in some range of small $\rho$ then the instability would exist however large $\chi$ were to grow. Therefore we identify the RG scale in this term with $\sqrt{\rho^2 + \chi^2}$ (this naturally happens in the D7 probe action where this quantity is the radial distance in the background space).

When  $\Delta m^2=0$, near the boundary which corresponds to the UV,  the solution  is given asymptotically  by  $\chi(\rho) = m + c/\rho^2$, with $c=\langle \bar{q}q \rangle$ of dimension three and $m$, the mass, of dimension one (note  $\chi$ and $\rho$ have dimension one).  For non-zero $\Delta m^2$, the solution  takes the form $L(\rho) = m \rho^{-\gamma} + c\rho^{\gamma-2}$, with 
\begin{equation} \Delta m^2 = \gamma (\gamma -2) \label{gammarel} \end{equation}
Here $\gamma$ is precisely the anomalous dimension of the quark mass.  
The BF bound below which an instability occurs is given by $\Delta m^2=-1$ when $\gamma=1$. To directly control the running of the dimension (which is our goal) it is best to allow $\Delta m^2$ to have $\rho$ dependence at the level of the equation of motion. This effectively neglects a term in the equation of motion  $\rho \chi^2 {\partial \over \partial L} \Delta m^2$ - this term would in any case only be large when $\Delta m^2$ is varying fast at the BF bound violation point.

We will fix the functional form of $\Delta m^{2}$ using the one loop running of the gauge coupling in QCD with  $N_{f}=3$ flavours transforming in the fundamental representation. This is found by solving:
\begin{equation}
{\cal Q} \frac{d \alpha}{d{\cal Q}}=-b_{0} \alpha^{2}, \hspace{1cm}
b_{0}=\frac{1}{6 \pi}\left(11 N_{c}-2 N_{f}\right) 
\end{equation}  with ${\cal Q}$ the renormalization group scale.

The one loop result for the anomalous dimension of the quark mass is
\begin{equation}
\gamma_{1}=\frac{3 C_{2}}{2 \pi} \alpha, \quad C_{2}=\frac{\left(N_{c}^{2}-1\right)}{2 N_{c}}
\end{equation}
We  stress that using the perturbative result outside the perturbative regime is a sensible but non-rigorous, phenomenological parametrization of the running.

We will identify the RG scale, ${\cal Q}$ with the AdS radial parameter $\sqrt{\rho^2 + \chi^2}$ in our model. Working perturbatively from the AdS result $m^{2}=\Delta(\Delta-4)$ we have
\begin{equation}  \label{qcddm}
\Delta m^{2}=-2 \gamma_{1}=-\frac{3\left(N_{c}^{2}-1\right)}{2 N_{c} \pi} \alpha
\end{equation}

To find numerical solutions for $\chi$'s vacuum configuration, we need an IR boundary condition. In top down models $\chi^{\prime}(0)=0$ is the condition for a regular solution \cite{Karch:2002sh}. In previous papers \cite{Alho:2013dka} using this model this condition has been replaced by the very similar on mass shell condition   $\chi\left(\rho=\chi_0\right)=\chi_0$ with $\chi^{\prime}\left(\chi_0\right)=0$. Here $\chi_0$ is the IR value of the quark mass where the on-shell condition $\rho=\chi$ is realized. Thus one shoots out form the 45$^{\circ}$ line in the $\chi-\rho$ plane to find the value of $\chi_0$ that gives the desired quark mass at some UV value - here we will require that the mass vanishes in the UV. The resulting solution for $\chi$ is shown in red in Figure 1. Note here and henceforth we will use $\chi_0$, the IR quark mass, to set the scales in the theory rather than $\Lambda_{QCD}$ in the one loop running but there remains just one scale introduced via $\Lambda_{QCD}$.

  In previous studies the solution above the on-mass shell point has been sufficient - bound states masses can be determined by looking at fluctuations of 
this solution. Now though we wish to compute the action of this configuration. There are two complications. 
Firstly since we imposed $\Delta m^2$ at the level of the equations of motion we have neglected one term dependent on the derivative of $\Delta m^2$ in the equation of motion so   it is inconsistent to then use $\Delta m^2$ directly in the action. Here though this is a small error since $\Delta m^2$ only has a large derivative in a very small region of $\rho$ and we will neglect this error. 

Secondly we have no solution below $\rho=\chi_0$ yet the chirally symmetric solution $\chi=0$, which we will want to
 
 \begin{center}
    \includegraphics[width=9cm]{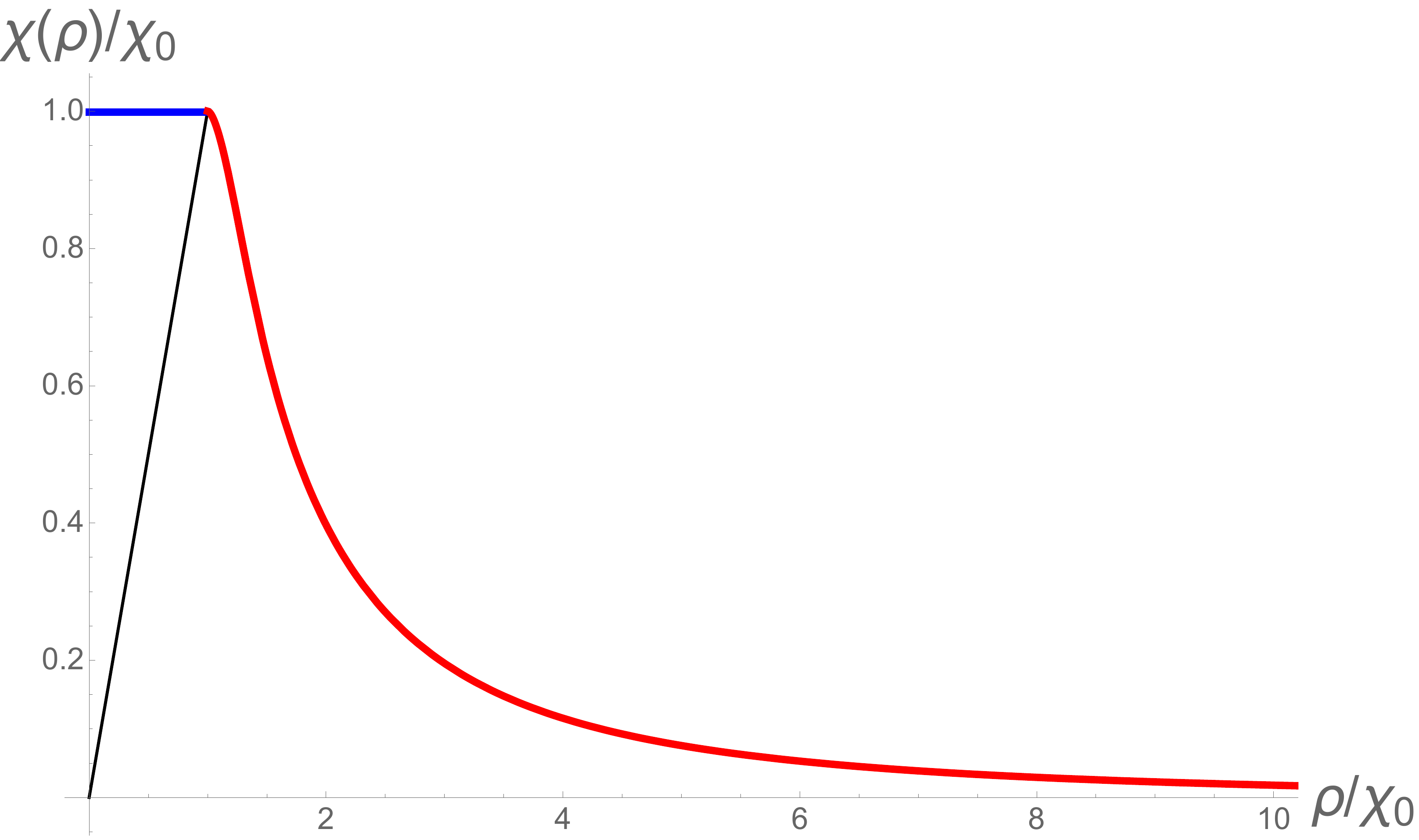}  \vspace{-0.4cm}
    
\noindent{{ \textit{Figure 1: The solution for the runing quark mass at $d=0$. The blue curve is the solution below the quarks' constituent mass with $\Delta m^2= 0$, and the red line is the solution above, where (\ref{qcddm}) holds. We also show the 45$^o$ line where we set boundary conditions on the solutions.}}}

\end{center} \vspace{-0.3cm}

 compare the action of our solution to, extends all the way to $\rho=0$. This problem will become worse below when we allow solutions with density where one expects the solution for $\chi$ to ``spike'' to the origin of the $\chi-\rho$ plane - with the current boundary conditions we will lose all of this part of the solution. Our resolution of this issue here is pragmatic, based on simply obtaining sensible looking solutions in the region interior to the 45$^\circ$ line. We will set $\Delta M^2=0$ in the region $\chi^2 + \rho^2 < 2 \chi_0^2$. The solutions of the equations of motion are then just those of the base D3/D7 probe system in (6). Thus for example at $d=0$ they are the solutions $\chi=m$. We will require the solution to match ($\chi$ and $\chi'$) to our exterior solution on the $\chi^2 + \rho^2 = 2\chi_0^2$ circle. Thus we extend the solution in Figure 1 into the IR with the blue solution shown. These solutions are now a sensible approximation to the forms found for $\chi$ in complete D3/D7 models with chiral symmetry breaking.

Even now there remains an ambiguity as to the constant prefactor between our UV and IR action pieces. We will keep this ambiguity as a multiplier $k_{IR}$ on the IR action. In the philosophy of this modelling we assume that chiral symmetry breaking occurs at a higher RG scale than confinement. Below $\chi_0$ the quarks should integrate out from the  dynamics leaving the pure glue theory to provide confinement. Since we don't include this dynamics our IR action is likely out by a constant factor. The $k_{IR}$ choice is one way to  include this factor in the dynamics. 

Thus the solution for $\chi$ in Figure 1 is our description of the $\mu=0$ vacuum of  QCD.  We will use the action of this configuration(for a given choice of $k_{IR}$)  to set our zero of potential energy. In fact this state will persist until quark density switches on 
at $\mu=\chi_0$ (a scale that is naturally of order 330 MeV in  QCD - one third of the proton mass). However, before that point we must allow for a density of nucleons to set in.  \vspace{-0.5cm}

\subsection{IIC Nuclear phase}\vspace{-0.5cm}

At small chemical potentials the nuclear transition in QCD is well understood: the confined, chirally broken vacuum is empty until a chemical potential of $\mu = 308.55$ MeV when there is a first order phase transition to nuclear matter. This transition is well studied and the nuclear matter EoS has been explored in \cite{Hebeler:2013nza}. There the authors combined observations of a $1.97$ solar mass neutron star with effective field theory (EFT) to construct the EoS,  extrapolating with a constrained piecewise polytropic form. Here holography is probably least able to help - given its origin at infinite $N_c$, baryons are naturally very heavy and far from the QCD limit. Thus, following several other authors \cite{{Hoyos:2016zke},{Ishii:2019gta},{Jokela:2018ers}}, we will simply use  the results of \cite{Hebeler:2013nza} to model the nuclear phase. Note there have been attempts to study the QCD nuclear phase 

\begin{center}
 \includegraphics[width=8cm]{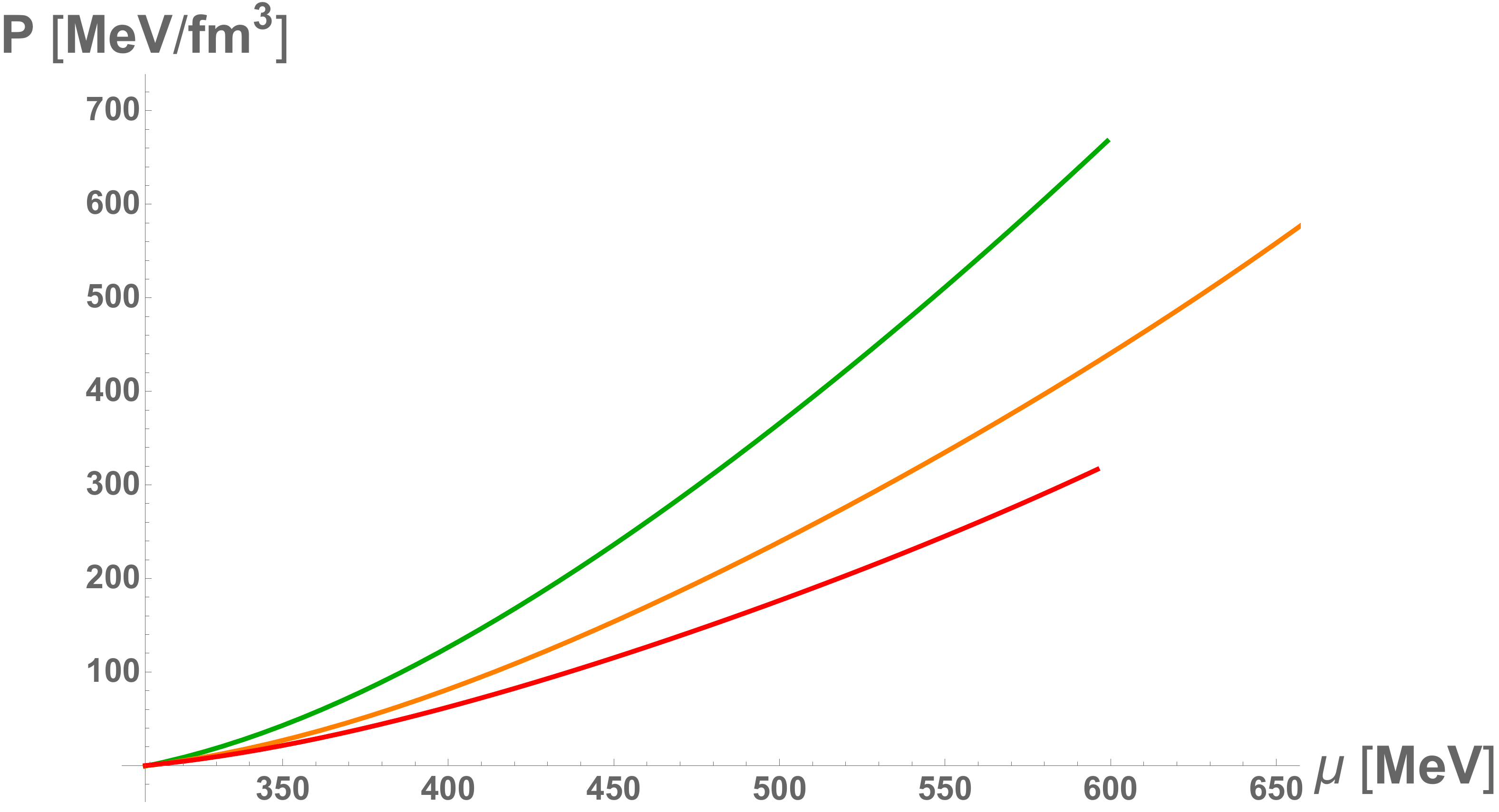}  
 
 \noindent{ \textit{Figure 2: Data for the nuclear phase taken from  \cite{Hebeler:2013nza}: we show  the pressure versus chemical potential.  The Green line represents a soft EoS, the orange a medium EoS and the red line a stiff EoS.}}
 
\end{center} 

\vspace{-0.3cm}

holographically, for example in \cite{Jokela:2018ers,Li:2015uea,Bergman:2007wp,Evans:2012cx}, but this will not be our focus in this paper. 

Three ansatz for the EoS (soft, medium and stiff) are presented in Table 5 of \cite{Hebeler:2013nza} - they give the energy density and pressure for different densities.  We have encoded their data as a Mathematica fitting polynomial for the analysis below and we plot these in Figure 2.

\subsection{IID The dense quark phases} \vspace{-0.5cm}

We next consider the (separate) transitions associated with the onset of quark density and to a chirally symmetric quark phase in our holographic model. We allow for a quark density by including a U(1) gauge field in addition to the action (11)
\begin{equation}
    \mathcal{L}=-\rho^3\sqrt{1+(\partial_{\rho}\chi)^2-(\partial_{\rho}A_t)^2}-\rho  \Delta m^2~ \chi^2
\end{equation} 

Here $A_t$ has UV asymptotic solution $\mu + d/\rho^2$ where $d$ is the density. We apply a Legendre transformation to obtain the action in terms of the density $d$
\begin{equation}
\Tilde{\mathcal{L}}=-\sqrt{\left(1+\chi'^2 \right) \left(\rho^6+d \right)}-\rho \chi^2 \Delta m^2
\end{equation}

\noindent Then the equations of motion are:
\begin{equation}
\begin{aligned}
\partial_\rho\left[\frac{\left(\rho^6+d \right)\partial_\rho \chi}{\sqrt{\left(1+\chi'^2 \right) \left(\rho^6+d \right)}} \right]- \rho \Delta m^2  \chi=0
\end{aligned}
\end{equation}
\begin{equation}\label{at}
(\partial_\rho A_t)^2=\frac{d^2\left(1+ \chi'^2 \right)}{\left(\rho^6+d \right)}
\end{equation}
Note in the first equation we have again suppressed the term $\rho  \chi^2 \frac{\partial}{\partial L} \Delta m^2$. 

We solve the equations of motion in two steps for each $d$.  We divide the space using the IR value of the quark

\begin{center}
    \includegraphics[width=9cm]{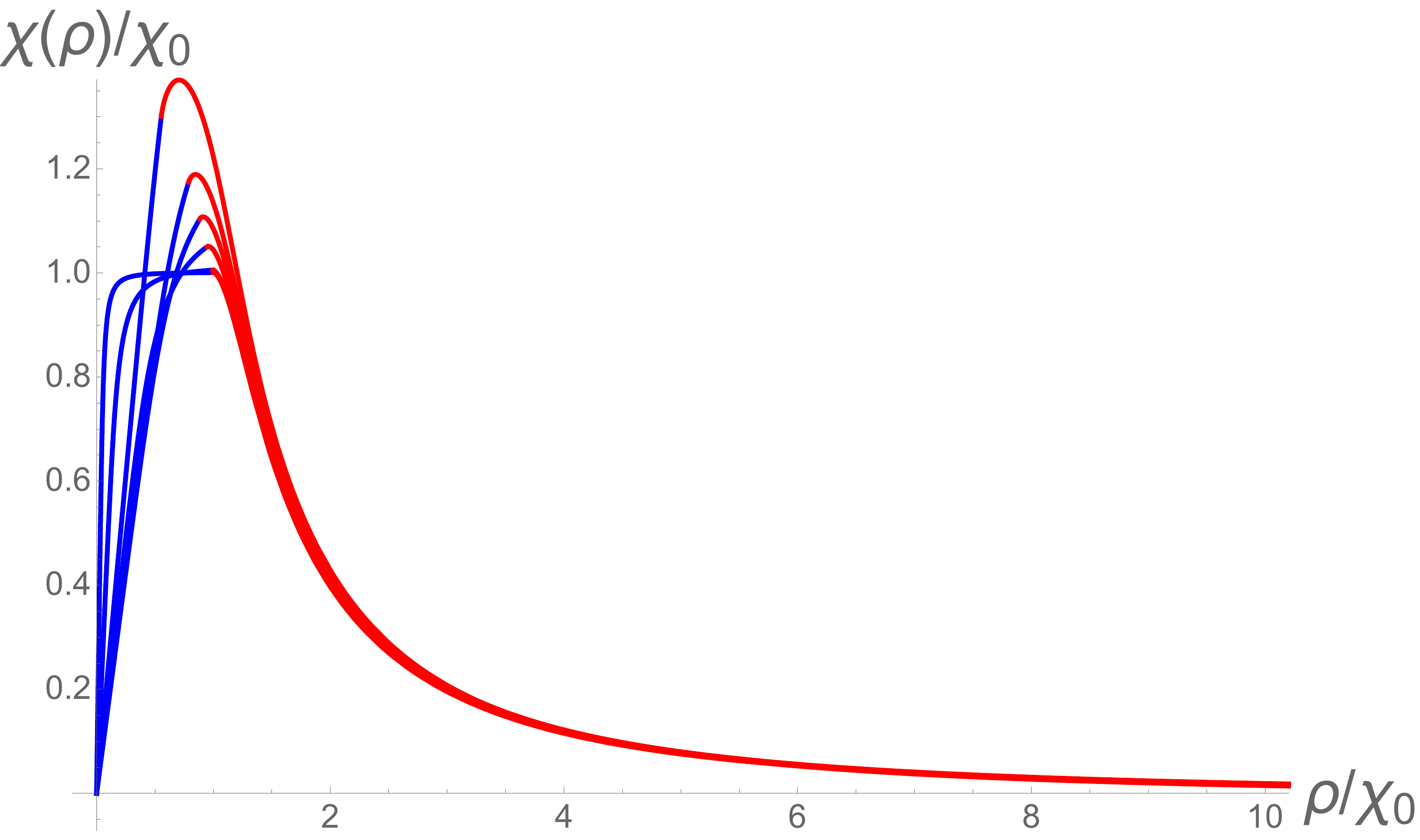}
    \includegraphics[width=9cm]{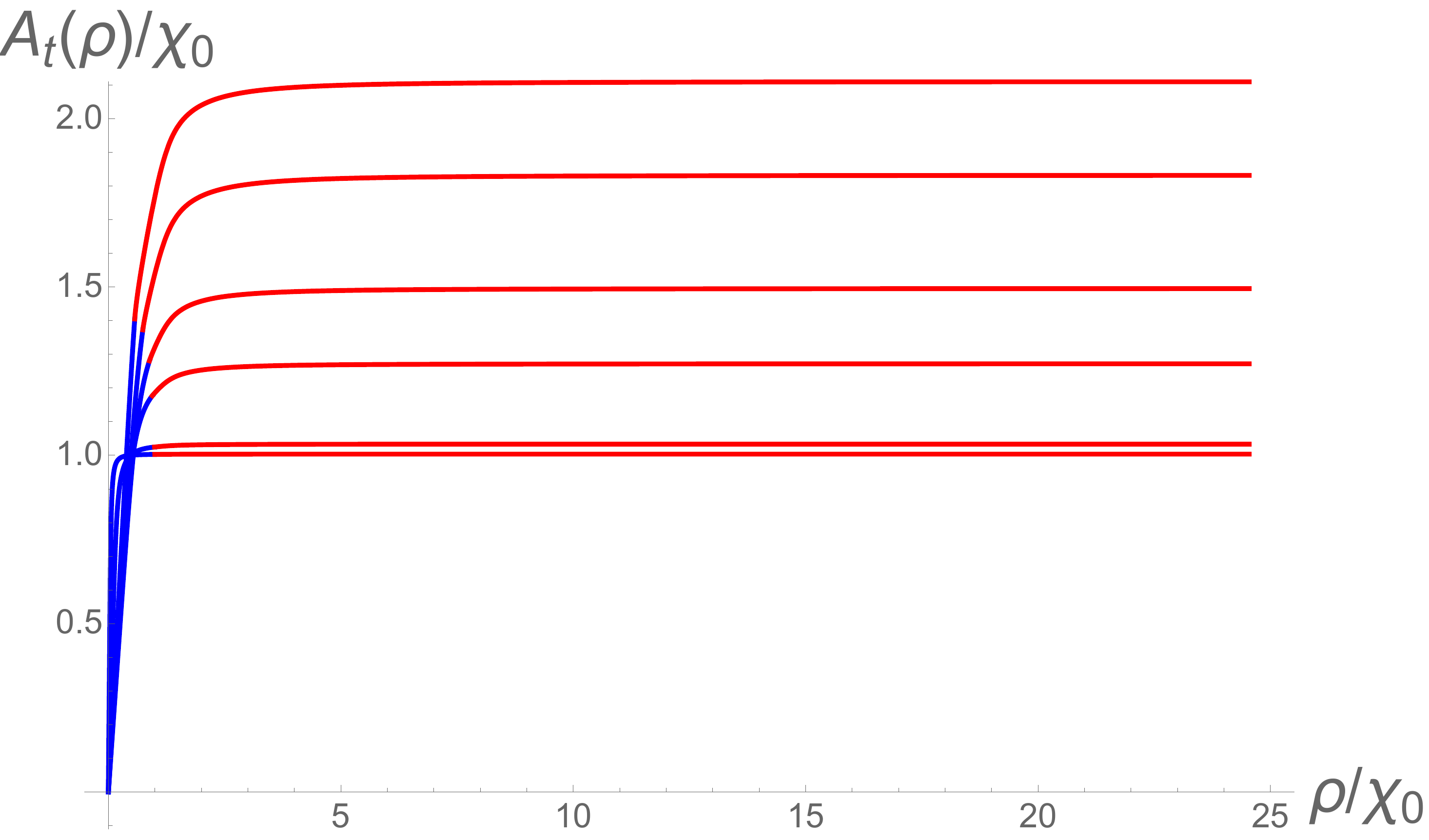}
   
\noindent{ \textit{Figure 3: Solutions for $\chi$ and $A_t$ when $d=0.00147\chi_0^3$, $0.0147\chi_0^3$, $0.147\chi_0^3$, $0.295\chi_0^3$, $0.488\chi_0^3$ and $0.554\chi_0^3$. The blue curve is the solution below the constituent mass scale and the red is that above.}} 

\end{center}  

mass at $d=0$, $\chi_0$. We obtain solutions for $\chi(\rho)$ and $A_t(\rho)$ in two intervals: first for  \mbox{$0\leq \rho^2 + \chi^2  \leq 2 \chi_0^2$}, which we call the region below the constituent mass and the other \mbox{$2 \chi_0^2 \leq \rho^2 + \chi^2 \leq \Lambda_{UV}^2$}, with $\Lambda_{UV}$ a large UV cut off, which we call the region above the constituent mass. 

Below the constituent mass we fixed $\Delta m^2$  to zero. For a given $d$ we shoot from the origin of the $\chi-\rho$ plane  with different gradients for $\chi$. We solve until we reach the surface \mbox{$\chi^2 + \rho^2 = 2\chi_0^2$} when we read off $\chi$ and $\chi'$ plus $A_t$.  These numerical solutions can also be checked against the 
analytic form in (6). External to the circle we use 
the running $\Delta m^2$ from the QCD perturbative running and match the initial conditions provided from the interior on the circle. 
We then seek amongst those solutions the one that shoots to a zero UV quark mass. Then in the UV we can read off the value of the chemical potential from the $A_t$ solution. We repeat this for each value of $d$.

The results are shown in Figure 3. The chirally broken phase exhibits a second order transition where density switches on. This behaviour is controlled by the low $\rho$ phase with $\Delta m^2=0$ - it is just the transition of the ${\cal N}=2$ model where a spike grows from the origin of the $\rho-\chi$ plane connecting to the flat embedding. The exterior region (in red) plays no role initially.  As $d$ increases the model resists returning towards the $\chi=0$ chirally symmetric phase with the maximum value of $\chi$ even increasing. This is the phase we call the deconfined massive quark phase.

After $d=0.554 \chi_0^3$ there are no non-trivial solutions that have a zero UV mass so by this value of $d$  a transformation to the chiral restored phase must have occurred (this puts some constraints on the parameter $k_{IR}$ as we will see).

We compute the free energy by obtaining the on-shell action for each value of $d$. The integration follows the same separation into the regions above and below the constituent mass.We weight the IR piece's action by the parameter $k_{IR}$.
One must be careful when splitting into sub-regions that any counter term is the same for each computation. We normalize so that the vacuum energy is zero for the $d=0$ embedding as previously discussed. 

We show some example plots of the pressure (minus the free energy) against $\mu$ for various $k_{IR}$ in Figure 4. For $k_{IR}=1$ and 2 the system does not make sense. The chirally broken state ceases to exist before it stops being the true vacuum. 
On the other hand for $k_{IR}=0.575$ the system is more 
sensible - the chirally restored vacuum becomes preferred 
and the chirally broken state becomes metastable before it ceases to exist. This provides a sensible description of a first order chiral restoration transition. We also show the case $k_{IR}=0.1$ where the transition occurs at lower $\mu$.

\begin{center}
    \includegraphics[width=9cm]{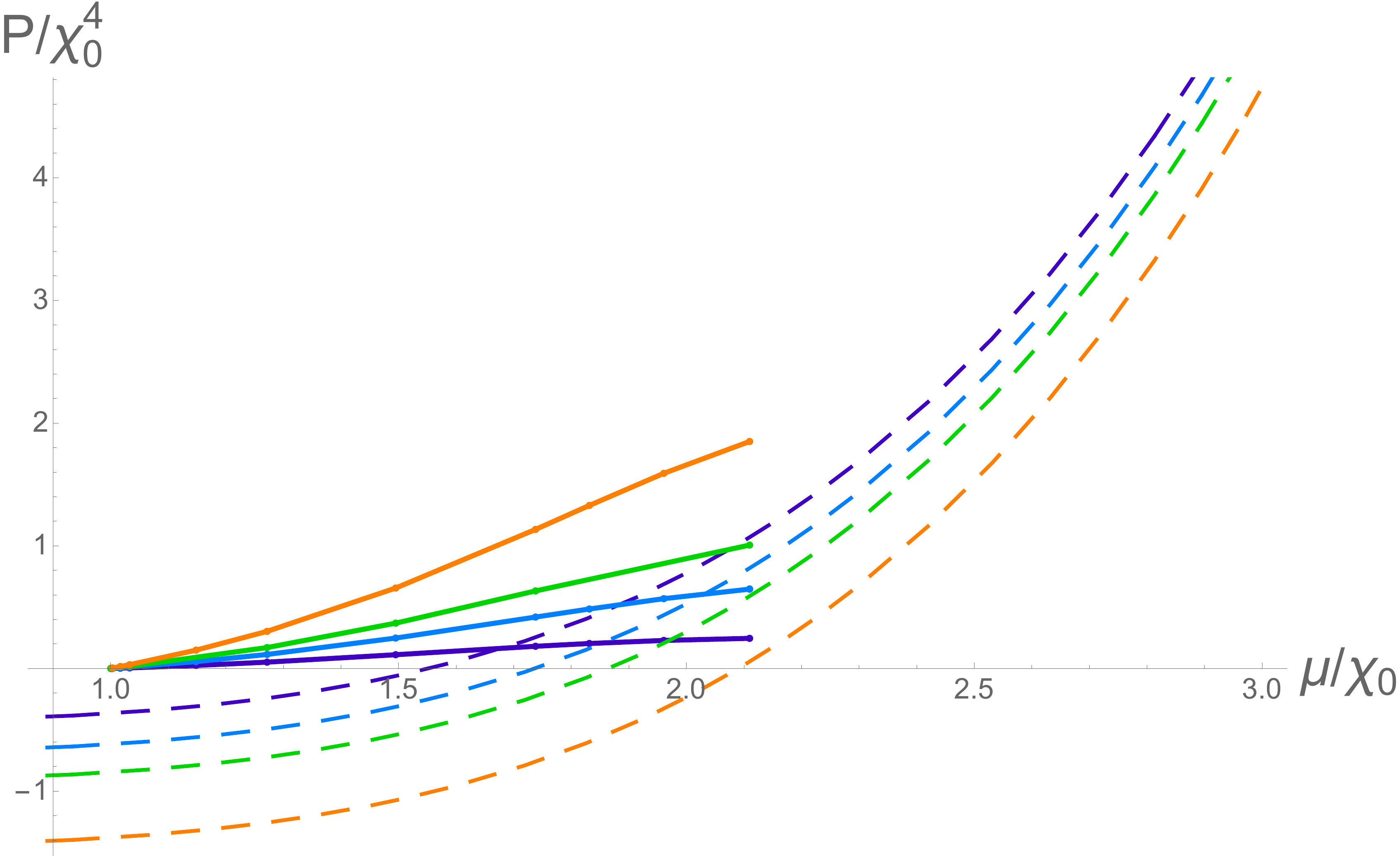}

    \noindent \textit{Figure 4: Pressure versus chemical potential. The solid line corresponds to the deconfined massive phase, and the dashed line represents the chirally restored phase ($\chi=0$ phase). The different colors represent different values of  $k_{IR}$;  $k_{IR}=0.1$ (purple), $k_{IR}=0.575$ (blue), $k_{IR}=1$ (green) and $k_{IR}=2$ (orange). }
\end{center}  \newpage

\begin{center}
    \includegraphics[width=9cm]{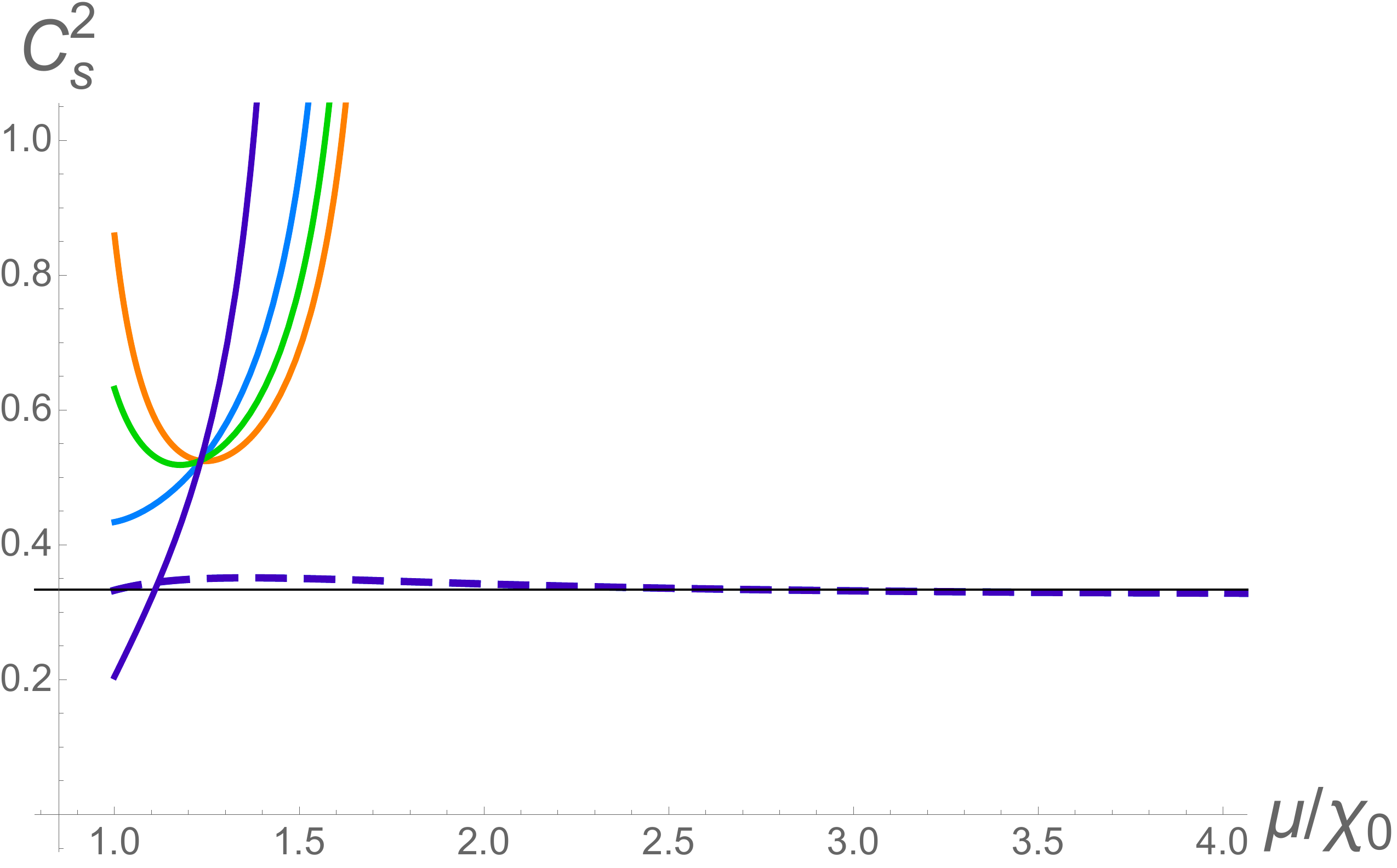}
    
\noindent{ \textit{Figure 5: The speed of sound squared as a function of $\mu$ for the Figure 4 solutions. The top lines represent the chirally broken phase with different values of  $k_{IR}$;  $k_{IR}=0.1$ (purple), $k_{IR}=0.575$ (blue), $k_{IR}=1$ (green) and $k_{IR}=2$ (orange).  The lower dark blue line corresponds to  the chirally restored phase ($L=0$  phase) which asymptotes to 1/3.  }}

\end{center}

Our expectation is that in the sensible systems the chirally broken phase is rather stiff. It is resisting the transition to the chirally restored phase. A good test of this is to determine the speed of sound squared, $c_s^2$ (which is
simply $\partial P / \partial {\mathcal E}$)- see Figure 5. We show the results for the four values of $k_{IR}$ in Figure 4 and also for the chirally symmetric phase $\chi=0$. We plot for values of $\mu$ above the transition where density switches on. The 
$c_s^2$ in the chirally symmetric phase is 1/3. The speed of sound in the chirally broken phase though rises much higher and even passes through the speed of light $c=1$. Note that for the cases of $k_{IR}=1,2$ the speed of sound has rather strange behaviour including a turning point - this suggests again that these choices of $k_{IR}$ do not make physical sense. The cases we will continue with between $k_{IR}=0.1-0.575$ have monotonic rising behaviour. For the moment we will allow the speed of sound to lie greater than one but will return to address this issue when we add a colour superconducting condensate to the chirally restored phase.

We can next set $\chi_0=330$ MeV and compare the free energy of these phases to the nuclear phase's free energy. We do this in Figure 6 (left) showing the cases $k_{IR}=0.1, 0.35$ and 0.575.  The transition to the nuclear phase occurs at 308 MeV. At 330 MeV the deconfined massive 
quark phase's pressure begins to rise. The $k_{IR}=0.575$ curve rapidly becomes the true vacuum relative to even the least stiff nuclear phase. The case $k_{IR}=0.1$ only becomes the true vacuum relative to the stiffest nuclear equation of state. For intermediate $k_{IR}$ one can achieve curves between these limits - for example $k_{IR}=0.35$ grows to dominate the medium and stiffest nuclear curves 

\newpage

\begin{center}
 \mbox{   \includegraphics[width=9cm]{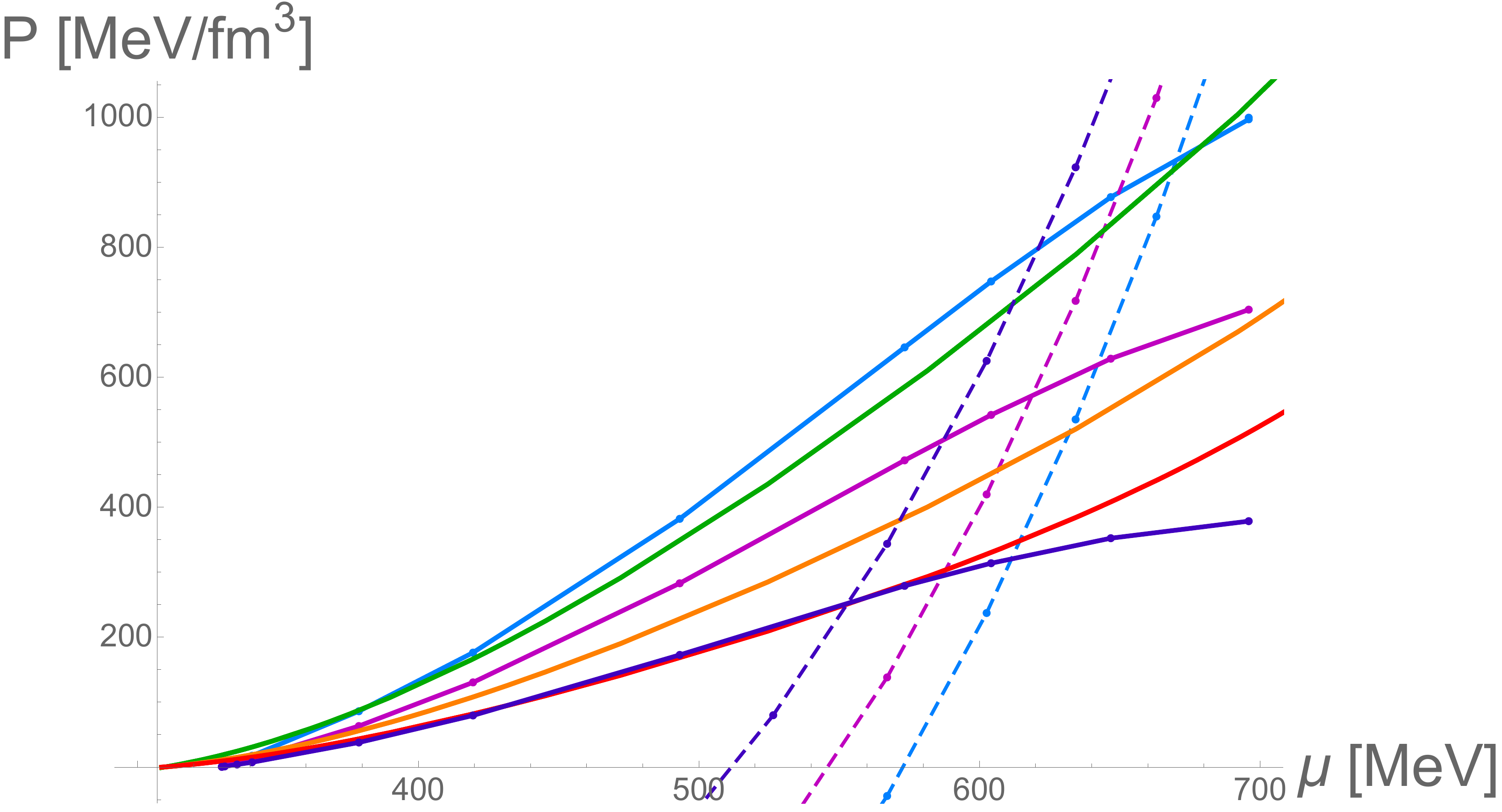}
     \includegraphics[width=9cm]{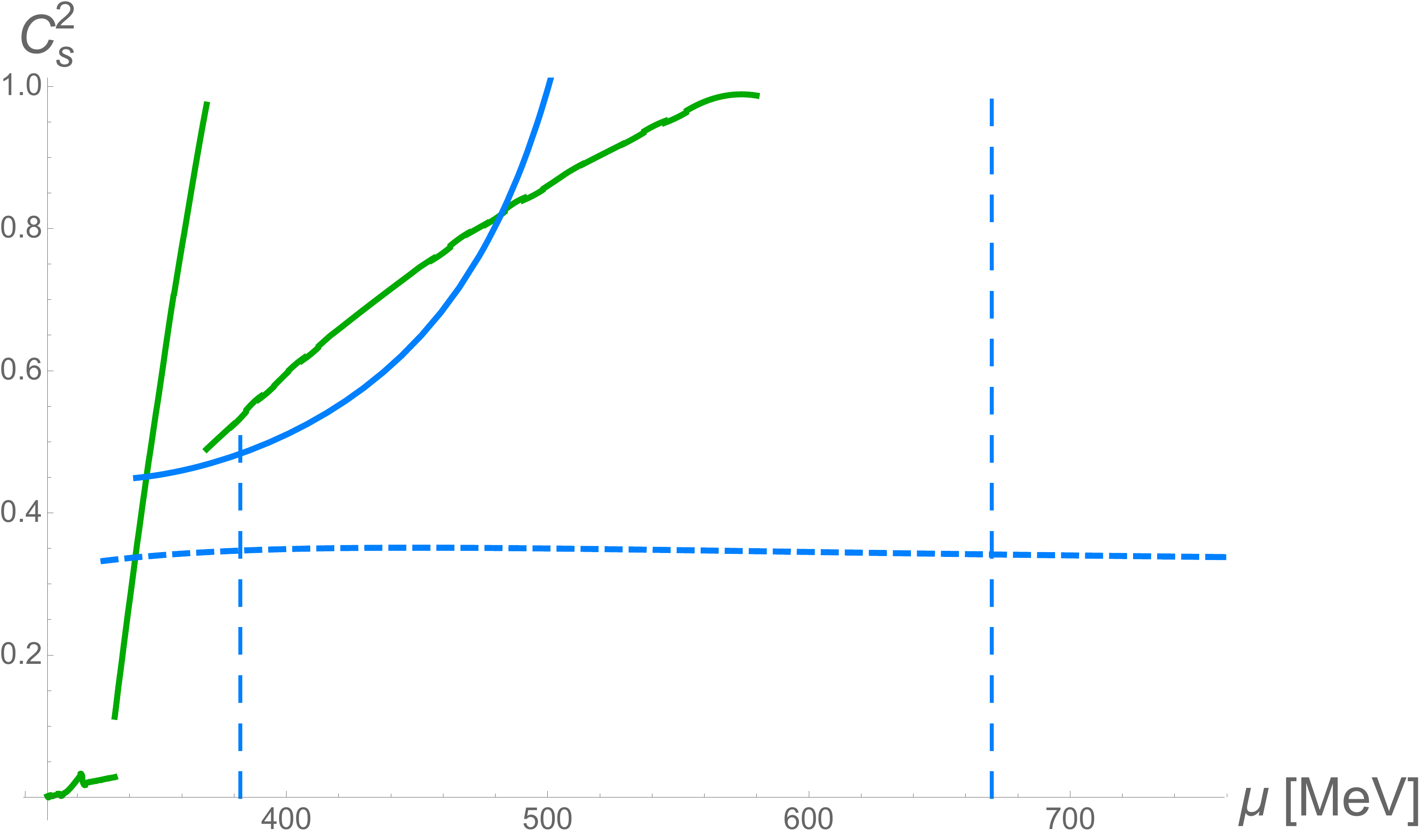} }
       
 \noindent{ \textit{ \mbox{Figure 6: Transitions from the nuclear phase to the deconfined phase: on the left a plot of pressure vs chemical  }
 \mbox{potential.The nuclear phase curves are copied from Figure 2. The other solid lines corresponds to the massive } \mbox{ chirally broken phase and dashed lines represent the chirally restored phase ($\chi=0$ phase), with $k_{IR}=0.575$ (blue),} \mbox{$k_{IR}=0.35$ (magenta) and with $k_{IR}=0.1$ (purple). For $k_{IR} > 0.575$  there is not a sensible  transition to the $\chi=0$  }
 \mbox{phase. For $k_{IR}<0.1$  the  nuclear phase  is always favoured, eg $k_{IR}=0.1$ only plays a role relative to the  stiff }
 \mbox{nuclear matter phase. On the right we show an example plot of the speed of sound squared vs chemical potential} \mbox{with colours corresponding to the left hand plot.The dashed vertical lines represent the transition from }
\mbox{nuclear to chirally broken quark matter and from chirally broken quark matter to the chirally restored phase.  }  \\  }}

\end{center}

but does not replace the soft  nuclear curve.

In the second plot in Figure 6 we show the variation of $c_s^2$  with $\mu$ for the case of the soft nuclear EoS (the EoS is piecewise constructed so there are discontinuities - we just quote these from \cite{Hebeler:2013nza}) and the $k_{IR}=0.575$ case for the chirally broken and chirally 
symmetric vacua. The vertical dotted lines show where the phase transitions between phases occur. For the moment $c_s^2$ rises above one before the final transition to the chirally restored phase. In the next section we will show that, by modifying the chirally restored phase by including a colour superconducting condensate, the transition 
away from the deconfined massive quark phase can occur earlier removing the region with $c_s^2>1$.

\subsection{IIE Colour superconducting phases}

There has been considerable speculation in recent years that there may be a colour superconducting phase of high density QCD \cite{Alford:2007xm}. In the presence of a Fermi surface and any attractive interaction the formation of a di-quark condensate is expected \cite{rgflow,Evans:1998ek}. In the two flavour theory the spinless condensate is in the fundamental representation of colour SU(3) and a single coloured $qq$ bilinear condenses. In the three flavour theory a colour flavour locking (CFL) state is expected to form with three $qq$ bilinears non-zero. 

Holographically it has been shown that the presence of a

  \newpage $\left. \right.$ \vspace{9.3cm}

chemical potential, through a dual gauge field $A_t$, causes a 
charged scalar's mass to be driven through the BF bound and cause condensation \cite{Evans:2001ab,Hartnoll:2008vx}. Thus baryon number charged operators such as the $qq$ bilinears would be expected to condense. The holographic dual is formally a description of gauge invariant operators and so it has proven hard to describe superconducting operators 
which  are colour charged and should break the gauge group. However, in 
\cite{BitaghsirFadafan:2018iqr}  we proposed that phenomenologically one can be more relaxed about this constraint. In a quark gluon plasma near a confining region of the phase diagram one expects a plasma of quarks but also potentially a plasma of colour magnetic monopoles that will play a part in the confinement mechanism. If these are present then the 
electric and magnetic gluon fields will all already have a Debye mass \cite{Freedman:1976xs} and the gauged nature of colour will be blurred. We proposed to simply neglect the back reaction of the coloured condensates on the gluons but use holography to describe the condensation mechanism and to compute the vacuum energy. In this spirit we will include the superconducting phase into our holographic model for neutron stars.

We will describe each condensing $qq$ operator by a scalar field $\psi_i$ that we introduce into the holographic model in analogy to the chiral condensate field $\chi$ (both are dimension 3 scalars). In addition though because the $qq$ operator carries baryon number it will couple directly to the baryon number U(1) gauge field in the bulk. Thus we propose the action
\begin{equation} \begin{array}{ccc}
\mathcal{L} & =& -\rho^3\sqrt{1+(\partial_{\rho}\chi)^2-( \partial_{\rho}A_t)^2} -\rho^3 g_{\rho \rho} \sum_i(D \psi_i)^2\\ &&\\ &&-\rho  \Delta m^2 \chi^2, \hspace{1cm} D_\mu = \partial_\mu  - i G[\rho] Q A_\mu\end{array}
\end{equation}
We have trialled actions where $\psi_i$ enter the square root term but have not been able to make them give sensible profiles for $\psi_i$ particularly because in the deep IR the square root approaches zero. Here the action for $\psi_i$ is the kinetic term emerging from the DBI action in the expansion where all fields are small and with the derivative   promoted to a covariant derivative.  This is intended in the same spirit as $\Delta m^2$ is added, being the leading term for $\chi$ when aspects of the metric or dilaton contribute to its running $\gamma$. 

$Q$ is the quark number charge of the $qq$ bilinear (which we set to 2). The final issue though is that we must match the running coupling strength, $G[\rho]$, of the U(1) gauge field. In principle one should match this as the coupling runs from the perturbative regime but it may not be  appropriate to just use the one loop running for $\alpha(\rho)$. In addition to that running one also expects this coupling to run logarithmically as one approaches the Fermi surface - see \cite{Evans:1998ek} for example. In addition the strength of the attraction in different channels depends on group theory factors which could suppress the Cooper pair condensation coupling by as a much as an order of magnitude \cite{BitaghsirFadafan:2018iqr}. Together these effects could, at larger $\mu$ change the coupling further. We will therefore take $G$ to have the form
\begin{equation} G[\rho]^2=\kappa~  \alpha(\rho)\end{equation}
where $\kappa$ is a free parameter we will vary. Note allowing this choice enables us to find a wider set of solutions than were found in \cite{Ghoroku:2019trx}.

We do not expect both $\chi$ and $\psi$ to condense together. For example, Lagrangian terms we could include such as $|\phi |^2 |\psi |^2$ would tend to fight against any BF bound violation for one field if the other field condenses. We will therefore just concentrate on Cooper pair formation in the chirally symmetric $\chi=0$ phase to see how its presence effects that phase. 

The equations of motion are
\begin{equation} \begin{array}{ccc}
\partial_\rho( \rho^3 \partial_\rho \psi_i ) + {G[\rho]^2 Q^2 \over \rho} A_t^2 \psi_i & = & 0 \\ &&\\
\partial_\rho( \rho^3 \partial_\rho A_t ) - \sum_i {G[\rho]^2 Q^2 \over \rho} \psi_i^2A_t & = & 0  \end{array}\end{equation}
We solve the equations between a large UV cut off where $\psi_i \sim J_i + O_i/\rho^2$ with $J_i$ a source and $O_i$ the Cooper pair vev, and the IR scale $\sqrt{2} \chi_0$ where the running has become strong enough to cause $\chi$ condensation at $\mu=0$. Now we need suitable IR boundary conditions. It is not clear what to pick although any none extreme choice give similar behaviour - we pick  $\psi_i' = - \psi/\chi_0$ which has the same proportionality as the usual holographic superconducting case where the embedding ends on a black hole.  Thus we can now set $\psi(\sqrt{2} \chi_0)$ to find solutions that asymptote to $J=0$.

For $A_t$ we use the ${\cal N}=2$ theory $A_t$ at $\chi=0$ for various $d$ and use the values of the solutions at $\rho=\sqrt{2} \chi_0$ to set boundary conditions for $A_t$ externally.

Finally we note that the number of $\psi_i$ fields is easily dealt with.  If there are $N$ such degenerate fields then in the lower equation of (22) there is simply a factor of $N$ - it can be absorbed into the normalization of $\psi_i$. Since the top equation in (22) is linear in $\psi_i$ this rescaling does not change the solution. Similarly at the level of the action (20) rescaling $\psi_i^2$ by $1/N$ whilst summing over $N$ copies leaves the action invariant. Thus the difference between the theory of the two flavour condensate (where there is one $\psi_i$) and the colour flavour locked phase (with three) is just a rescaling of the condensate by a factor of $\sqrt{3}$.  We will therefore restrict to one $\psi_i$ for the analysis to come but the free energy/pressure analysis is the same for the colour flavour locked phase. 

The proof of all this construction is whether we obtain sensible phenomenology for the Cooper pair formation.

 In Figure 7  we plot some example embeddings for $\kappa =1$ and varying $\mu$. We indeed find profiles that asymptote to $J=0$ at each $\mu$ and where the gap size grows with $\mu$. We plot $O$ against $\mu$ for varying $\kappa$ in Figure 8. Now we can see that there is a second order transition (from the chirally symmetric vacuum) to the colour superconducting phase with $\mu$.  Note that this will not be a physical transition because at lower scales the  deconfined massive phase is preferred to the $\chi=0$ state - the true transition will be a first order transition from the deconfined massive phase to the superconducting phase.
 For $\kappa=10$, \vspace{-0.5cm}

\begin{center}
    \includegraphics[width=9cm]{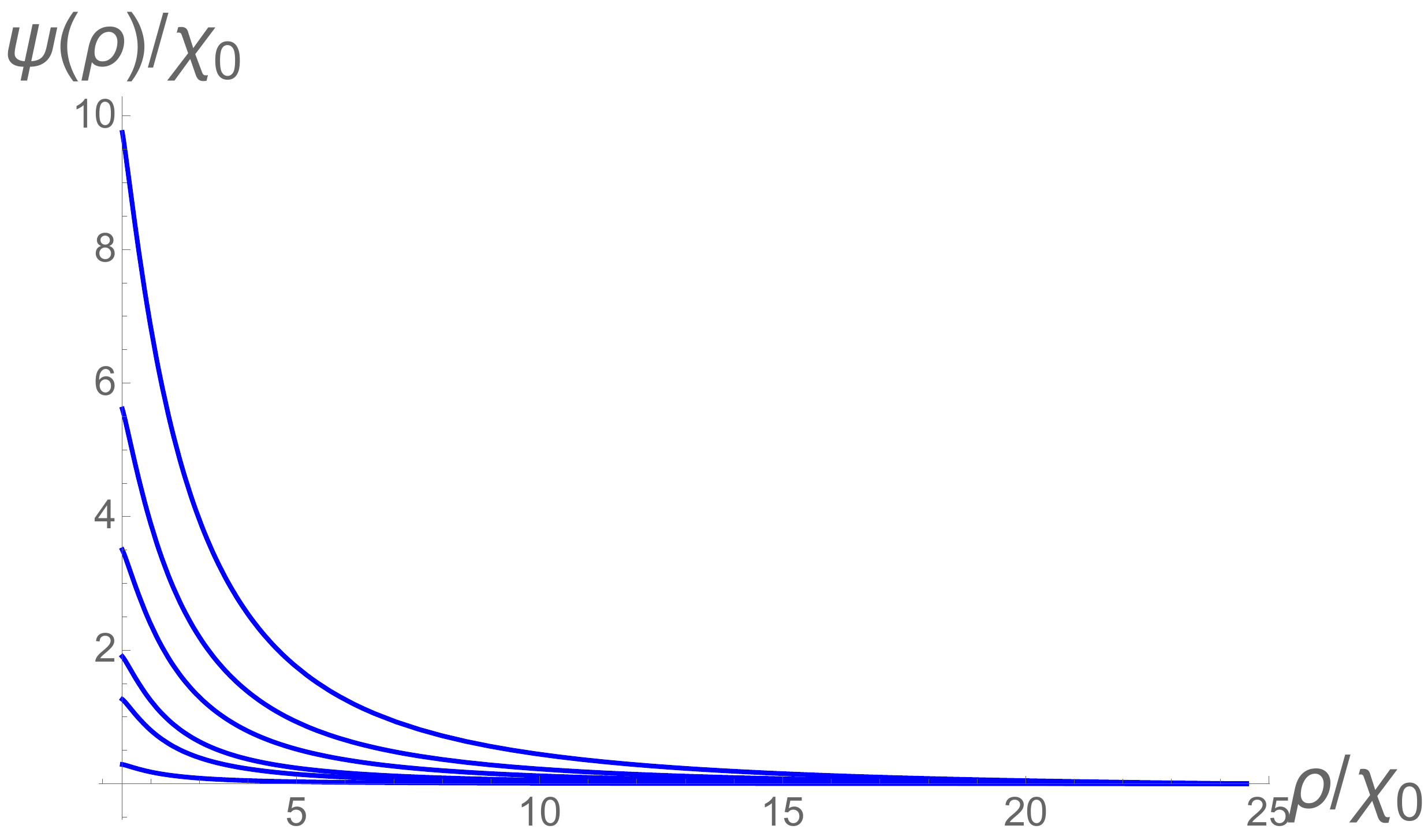}

\noindent  \textit{Figure 7. Solutions for $\psi$  in units of $\chi_0$ for $\kappa =1$ and different values of $\mu$ after condensation is triggered. }
\end{center} \vspace{-0.5cm}

\begin{center}
    \includegraphics[width=9cm]{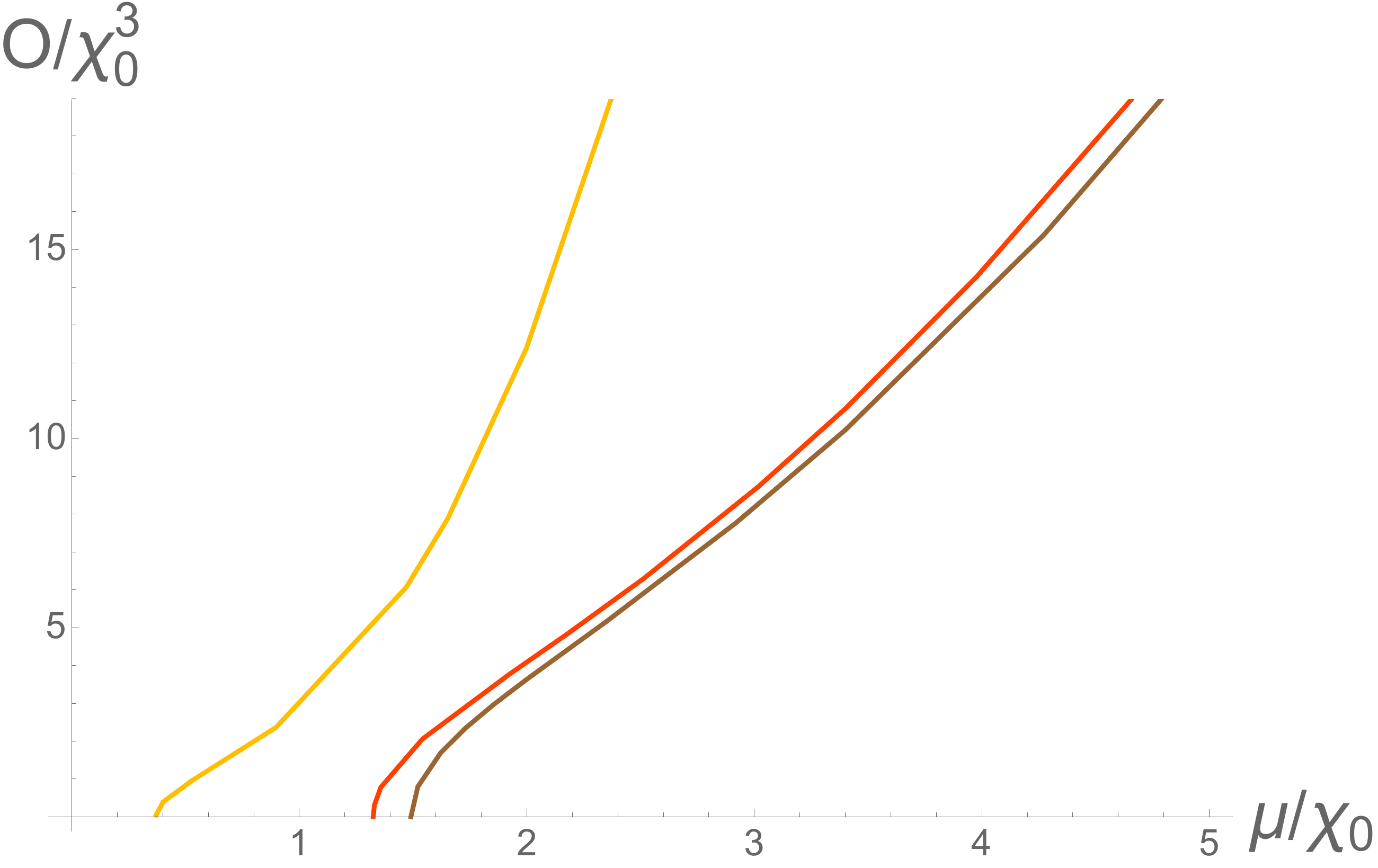}

 \noindent \textit{Figure 8. Cooper pair condensate as a function of the chemical potential for different values of $\kappa$; $\kappa=10$ (yellow), $\kappa=1$ (red) and $\kappa=0.85$ (brown).}
\end{center}

\begin{center}
    \includegraphics[width=8cm]{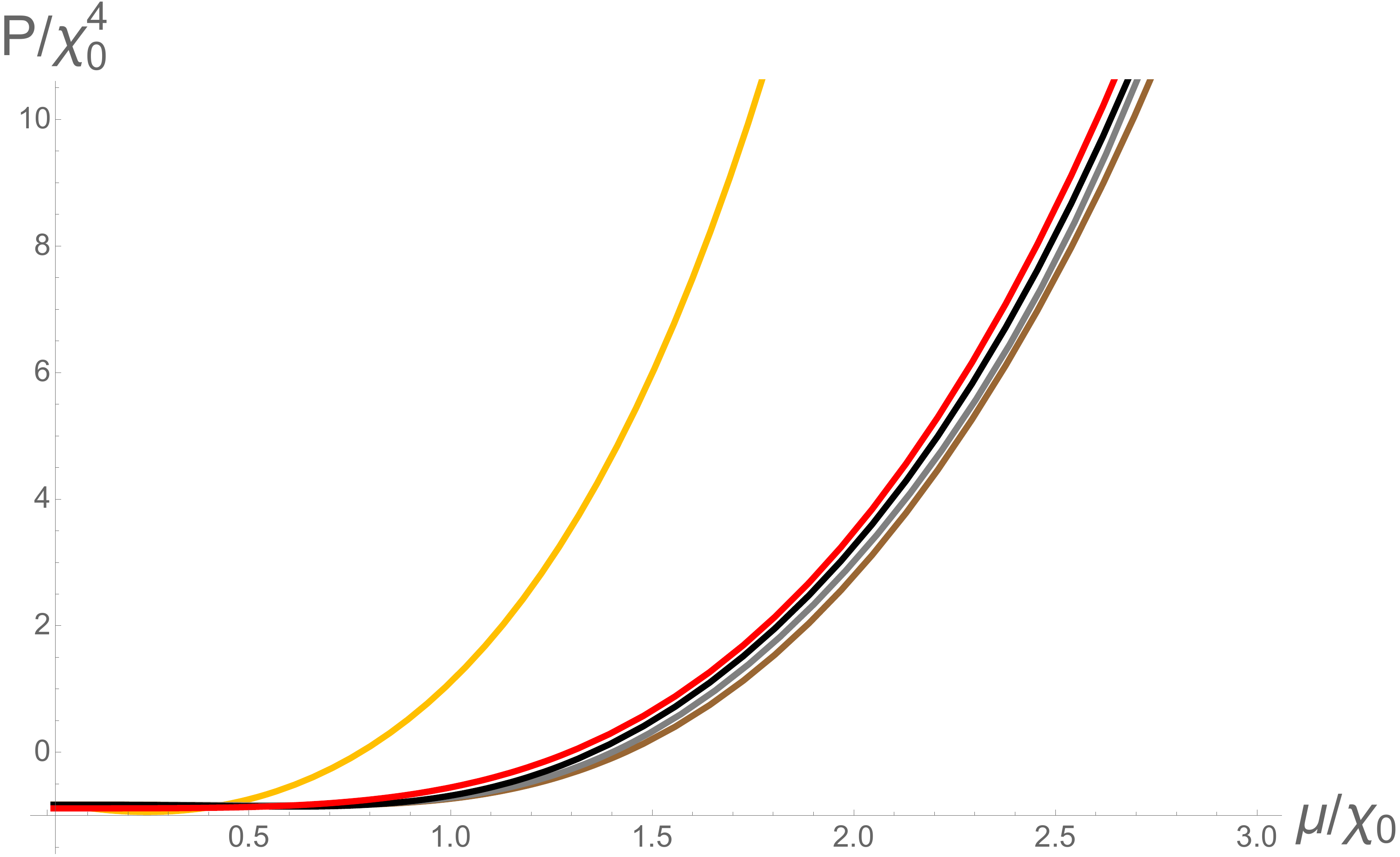}
 
  \noindent  \textit{Figure 9. Pressure versus chemical potential in the superconducting vacuum for different values of $\kappa$; $\kappa=10$ (yellow), $\kappa=1$ (red), $\kappa=0.94$ (black), $\kappa=0.89$ (gray) and $\kappa=0.85$ (brown).}
\end{center}

at  $\mu \simeq 1.5 \chi_0$ (approximately 500 MeV) where this first 
order transition to this phase will occur, the   condensate's scale is of order $\chi_0^3$ - (330 MeV)$^3$ 
- which is possibly large relative to the supposed gap scale although it serves as a sensible upper possible case. For $\kappa=1$ the condensate only 
switches  on close to $\mu \simeq 1.5 \chi_0$ and can be an order of magnitude smaller which is again a sensible lower estimate for 
the condensate's value. As $\mu$ increases in all cases the condensate grows in rough proportion to $\mu$.

We plot the pressure (minus the free energy) of the solutions in Figure 9. We see that the pressure of the phase is raised depending on the size of $\kappa$. We will adjust $\kappa$ to move the transition to the deconfined phase (now with colour superconductivity)  of Figure 6 to lower $\mu$ so that the speed of sound in the  deconfined  massive quark phase never rises above one.
We can make the transition occur just before the speed of sound passes through 1 with $\kappa=0.85$ for the case of $k_{IR}=0.575$; $\kappa=0.89$ for $k_{IR}=0.35$ and finally $\kappa=0.94$ for the case of $k_{IR}=0.1$
 
 To display this graphically we again set $\chi_0=330$MeV and compare the free energy of these phases to the
 nuclear phases' free energy and to the   deconfined massive quark phase of the previous sections. We do this in Figure 10 (top left) showing the cases $k_{IR}=0.1, 0.35$ and 0.575. 
As seen before the transition to the nuclear phase occurs at 308 MeV. At 330 MeV the deconfined massive 
quark phase's pressure begins to rise. The $k_{IR}=0.575$ curve rapidly becomes the true vacuum relative to even the least stiff nuclear phase. The case $k_{IR}=0.1$ only 
becomes the true vacuum relative to the stiffest nuclear    equation of state. For intermediate $k_{IR}$ one can achieve curves between these limits - for example $k_{IR}=0.35$ grows to dominate the medium and stiffest nuclear curves but does not replace the soft  nuclear curve.

Now though we also include the chirally restored vacuum with colour superconductivity curves for  $\kappa=0.94, 0.89$ and $0.85$. They rise sharply in pressure and become the true vacuum in the range $\mu=$450-500 MeV.

In the remaining plots in Figure 10 we show again the variation of $c_s^2$ with $\mu$ in a number of these scenarios.

      For example, in the upper right we show $c_s^2$ for the soft nuclear EoS, for the $k_{IR} =0.575$ case for the chirally broken, and for the $\kappa=0.85$ case of the colour superconducting vacuum. The vertical dotted lines show where the phase transitions between phases occur. We have tuned $\kappa$ so that the transition to the colour superconducting phase occurs just before the speed of sound passes through 1 (in order to have the stiffest EoS we can). The inclusion of colour superconductivity in the chirally restored phase does raise $c_s^2$ but only a little to around 0.4$c^2$. This will not be sufficient to support neutron stars if this material forms the core as we will see in the next section. The crucial role colour superconductivity is playing here is to reduce the critical $\mu$ for the transition from the deconfined  massive quark phase to ensure $c_s^2$ doesn't rise above 1.
      
      The lower two figures in Figure 10 show a variety of scenarios for the medium and stiff nuclear EoS.  In all cases there are, with increasing $\mu$, the phases: chiral broken; nuclear; massive  deconfined quark; chirally restored with colour superconductivity. In each case $c_s^2$ rises close to 0.7-0.8 in the nuclear
phase then to close to 1 in the deconfined massive  quark phase.    \vspace{-0.5cm}

\begin{center}
 \mbox{   \includegraphics[width=9cm]{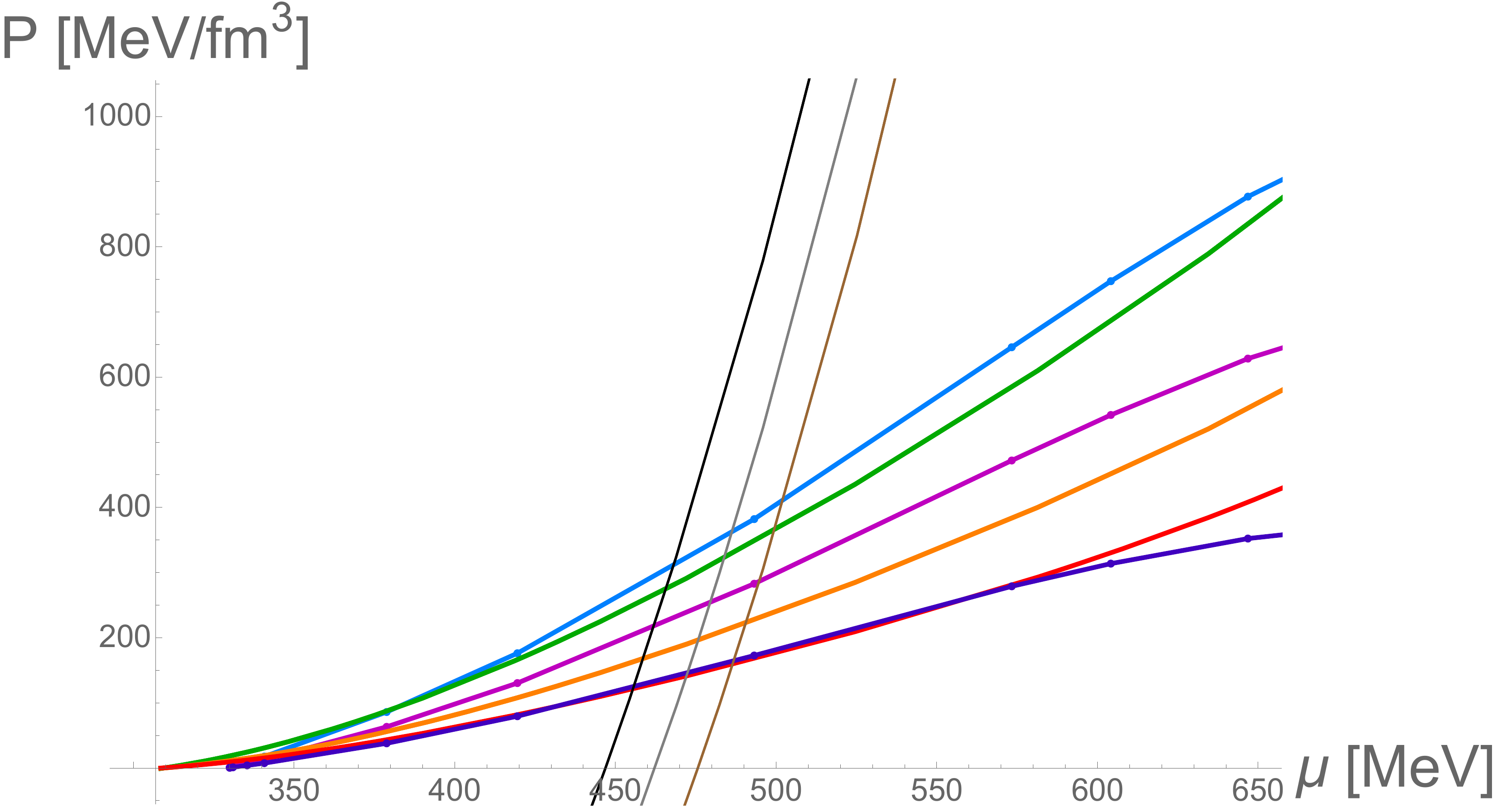}
     \includegraphics[width=9cm]{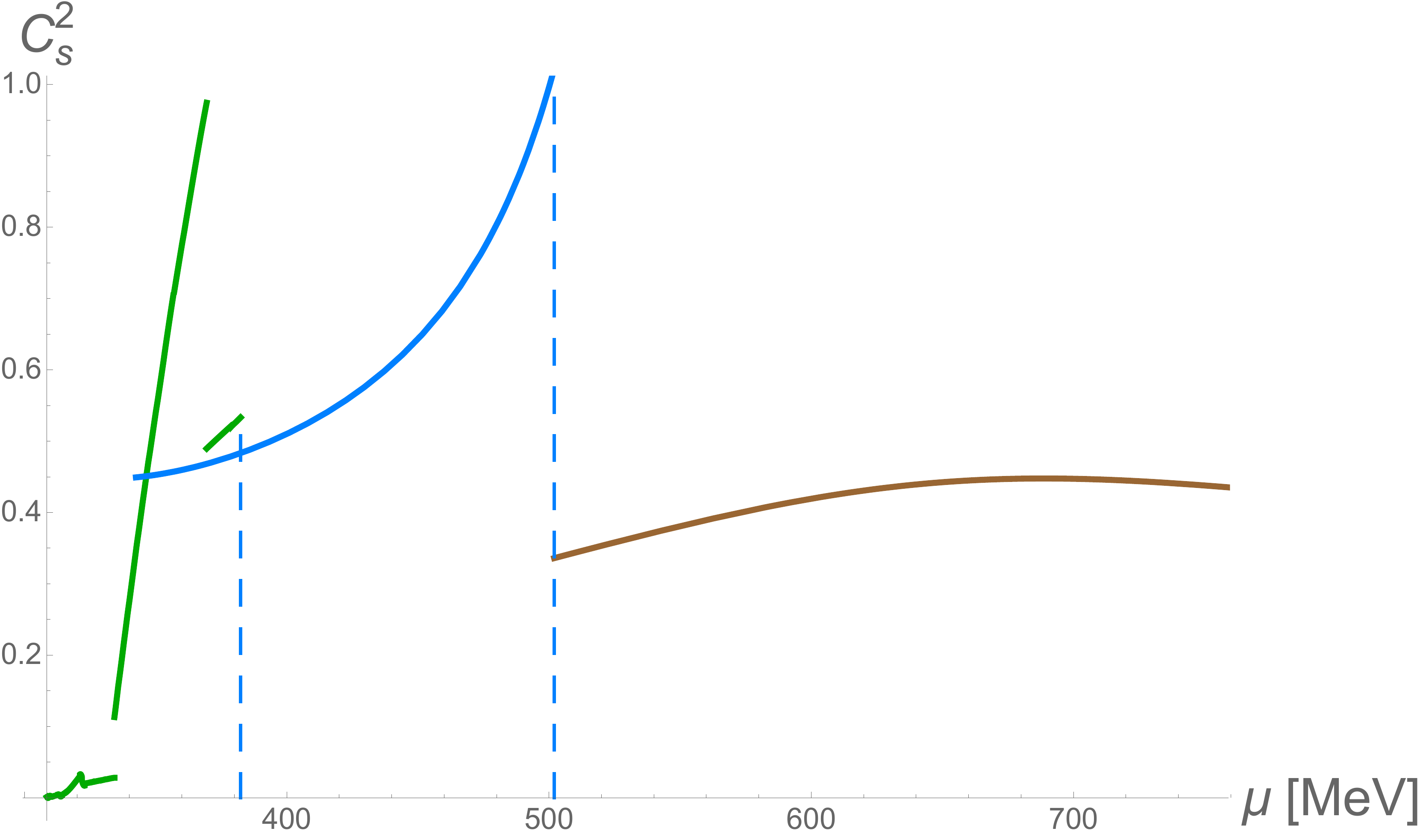} }
 \mbox{     \includegraphics[width=9cm]{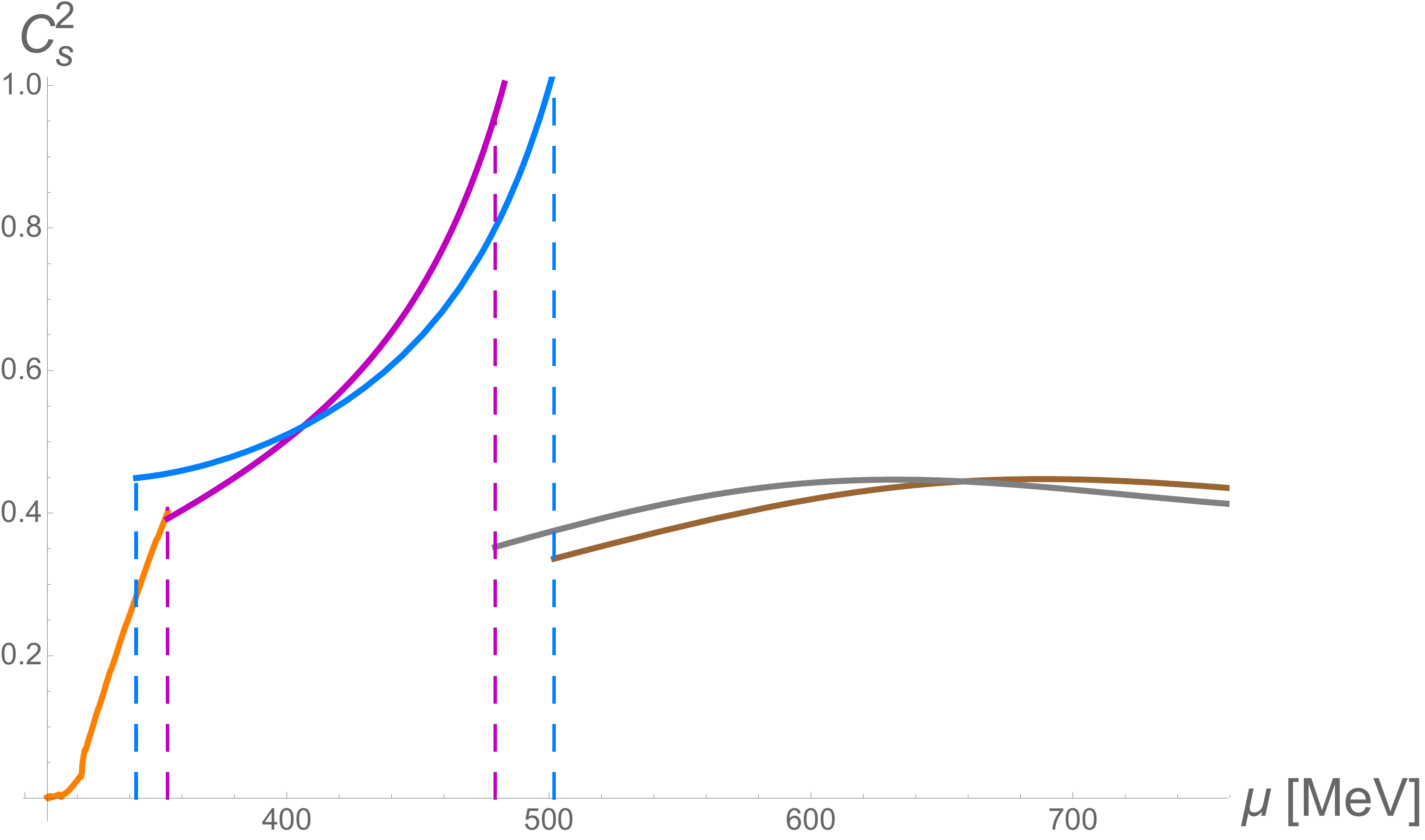}
       \includegraphics[width=9cm]{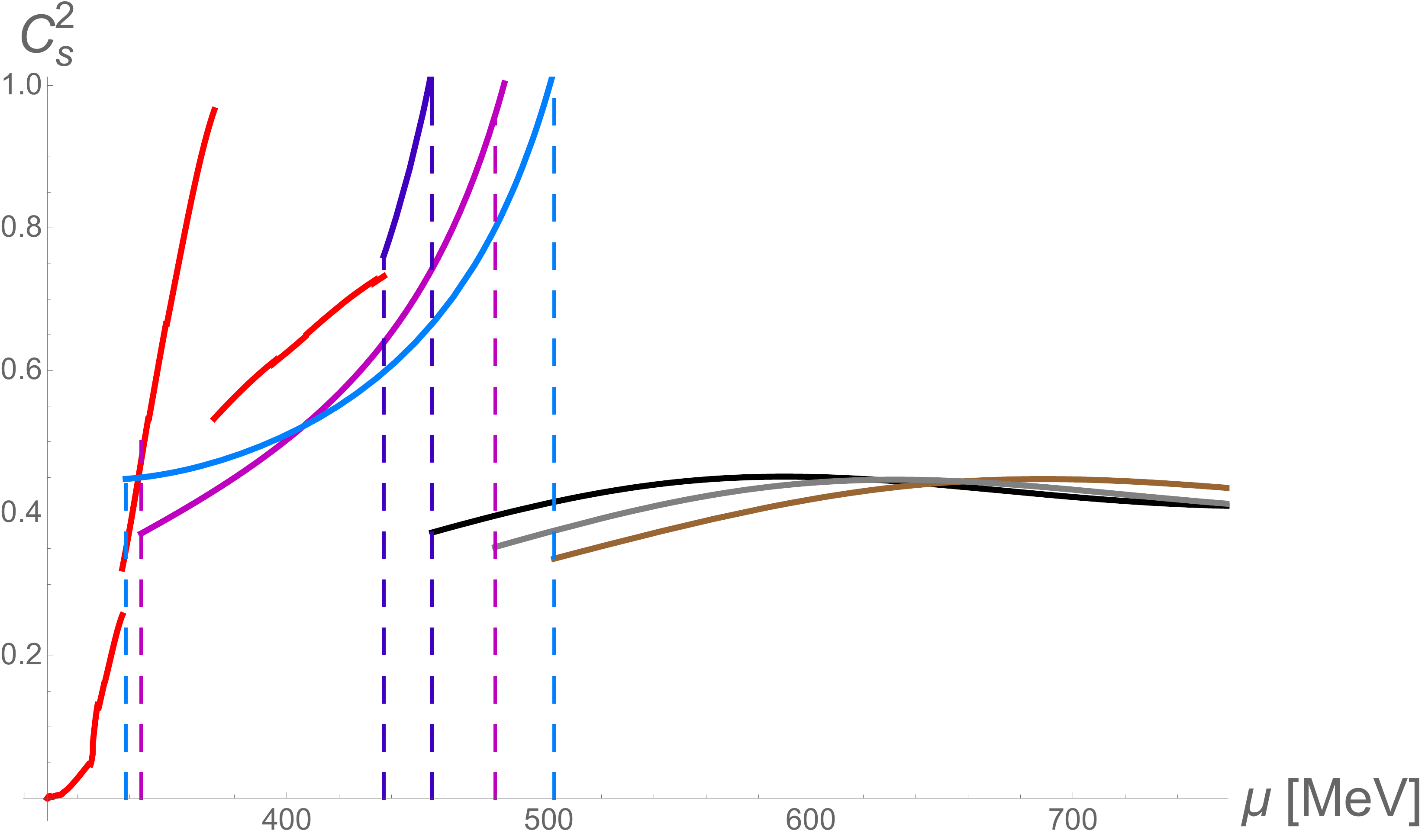}}
       
 \noindent{ \textit{ \mbox{Figure 10: Transitions from the nuclear phase to the deconfined phase and then to superconducting phase. The }  \\
 \mbox{different colors represent the cases of massive chirally broken phase with $k_{IR}=0.575$ (blue), $k_{IR}=0.35$ (magenta)   }   \\
 \mbox{and with $k_{IR}=0.1$ (purple) as in Figure 6, whereas the superconducting cases are $\kappa=0.85$ (brown),  $\kappa=0.89$ (gray)  }
 \mbox{and  $\kappa=0.94$ (black). The dashed  vertical lines represent the transition from nuclear to chirally broken quark matter}\\
 \mbox{ and from chirally broken quark matter to the superconducting phase.}
}}
\end{center}

\section{III Neutron star mass-radius relations}
The mass-radius relation for neutron stars is determined by the EoS of the neutron/quark matter. One solves the Tolman-Oppenheimer-Volkov (TOV) equations (see for example \cite{Haensel_Neutron_Stars1,compactstarbook})
\begin{eqnarray}
\frac{dP}{dr}&=&-G_N\left( \mathcal{E}+ P\right)\frac{M+ 4\pi r^3P}{r(r-2G_N M)},\\
\frac{dM}{dr}&=&4\pi r^2\mathcal{E} \qquad
\end{eqnarray} 
which are the relativistic equations that model hydrostatic equilibrium inside the stars. $G_N$ is Newton's constant. Here $M$ and $P$ are the mass and pressure in the star as a function of radius $r$. To integrate the equations we need to input the EoS $\mathcal{E}(P)$, as well as the central pressure $P_c=P(r = 0)$ as initial condition, and the output are the mass $M(r)$ and pressure $P(r)$ of the corresponding star. The radius $R$ of the star will be the value of $r$ at which the pressure vanishes.  Then varying
  
   \newpage  $\left. \right.$ \vspace{13.3cm}

 the initial condition $P_c$ as a parameter we can construct a curve for the mass of the star $M_\odot=M(r=R)$ against $R$.

It is useful to place the TOV equations in their dimensionless form:
\begin{eqnarray}
\frac{dp}{d\xi}&=&-B\frac{ye\left(1+\frac{p_0}{\epsilon_0}\frac{p}{e}\right)}{\xi^2(1-2B\frac{p_0}{\epsilon_0}\frac{y}{\xi})}\left(1+A\frac{p_0}{\epsilon_0}\xi^3\frac{p}{y}\right),\\
\frac{dy}{d\xi}&=&A\xi^2e(\xi) \qquad\qquad\qquad
\end{eqnarray} 
Where $r=r_0\xi$, $M=m_0y(\xi)$, $P=p_0p(\xi)$, $\mathcal{E}=\epsilon_0e(\xi)$, $A=\frac{4\pi r^3_0 \epsilon_0}{m_0}$ and  $B=\frac{Gm_0\epsilon_0}{p_0r_0}$. 

For the numerics one can fix the scale with, for example, the value of $p_0=\epsilon_0=\frac{(308.55 MeV)^4}{\pi^2}$ as is sensible in the context of the nuclear equation of state discussed above; this choice then fixes the rest of our scale parameters.

One can make a radial perturbation about a solution. In terms of  the mass vs radius curve one increases the value of the central density $\mathcal{E}_c$ whilst keeping the same mass. If $\frac{\partial M_\odot(\mathcal{E}_c)}{\partial\mathcal{E}_c} >0$ then the corresponding equilibrium solution for this new configuration has a higher mass and  therefore there is a deficit of mass. The gravitational force then needs to be balanced by increasing the central pressure. The forces acting on the matter in the star will therefore act to return the new configuration toward its original unperturbed state. However for the case in which $\frac{\partial M_\odot(\mathcal{E}_c)}{\partial\mathcal{E}_c} \leq 0$, if the star is perturbed, the forces acting on the perturbed star will act to drive it further from its original point in the mass vs radius curve. Therefore the  condition for stability is given by  
\begin{equation} \label{stability}
    \frac{\partial M_\odot(\mathcal{E}_c)}{\partial\mathcal{E}_c} >0.
\end{equation}
As mentioned in \cite{Alford:2017vca} we can also determine the stability of a star from the mass vs radius curve using the Bardeen, Thorne and Meltzer (BTM) criteria \cite{BardeenThorneMeltzer} which established a simple formulation to know if all its radial modes are stable: \vspace{-0.3cm}

\begin{enumerate}
\item At each extremum where the $M_\odot(R)$ curve  rotates counter-clockwise with increasing central pressure, one radial stable mode becomes unstable.
\item  At each extremum where the $M_\odot(R)$ curve  rotates clockwise with increasing central pressure, one unstable radial stable mode becomes stable.
\end{enumerate} \vspace{-0.3cm}

We now perform these calculations for the EoS we obtained in Figure 10. To summarize our model has three parameters: $\chi_0$ which is the IR quark mass that we have set to 330MeV; $k_{IR}$ which must lie below 0.575 to ensure there is a sensible transition from the deconfined massive quark phase to the chirally restored phase; and $\kappa$ which determines the strength of the superconducting interaction which we have used to move the chiral restoration transition so that $c_s^2$ is never greater than one.

We present the mass radius relations for neutron stars that we obtain in Figure 11.   At low mass the stars are entirely neutron matter and the $M-R$ plot is determined by the nuclear EoS - these are the green, orange and red lines depending on the choice of nuclear EoS. In each case though we now propose a transition to hybrid stars with deconfined massive  quarks in the core. These are the blue, magneta or purple lines departing from the nuclear curves - the stable parts of these curves are marked by dashed lines. Finally there is a transition in the centre of the star to chirally symmetric quark matter leading to unstable stars. These branches angle off the deconfined massive  curves down to the left - the higher example of this branch in each case is the chirally restored phase without superconductivity whilst the lower example has a superconducting condensate present.

\begin{center}
    \includegraphics[width=9cm]{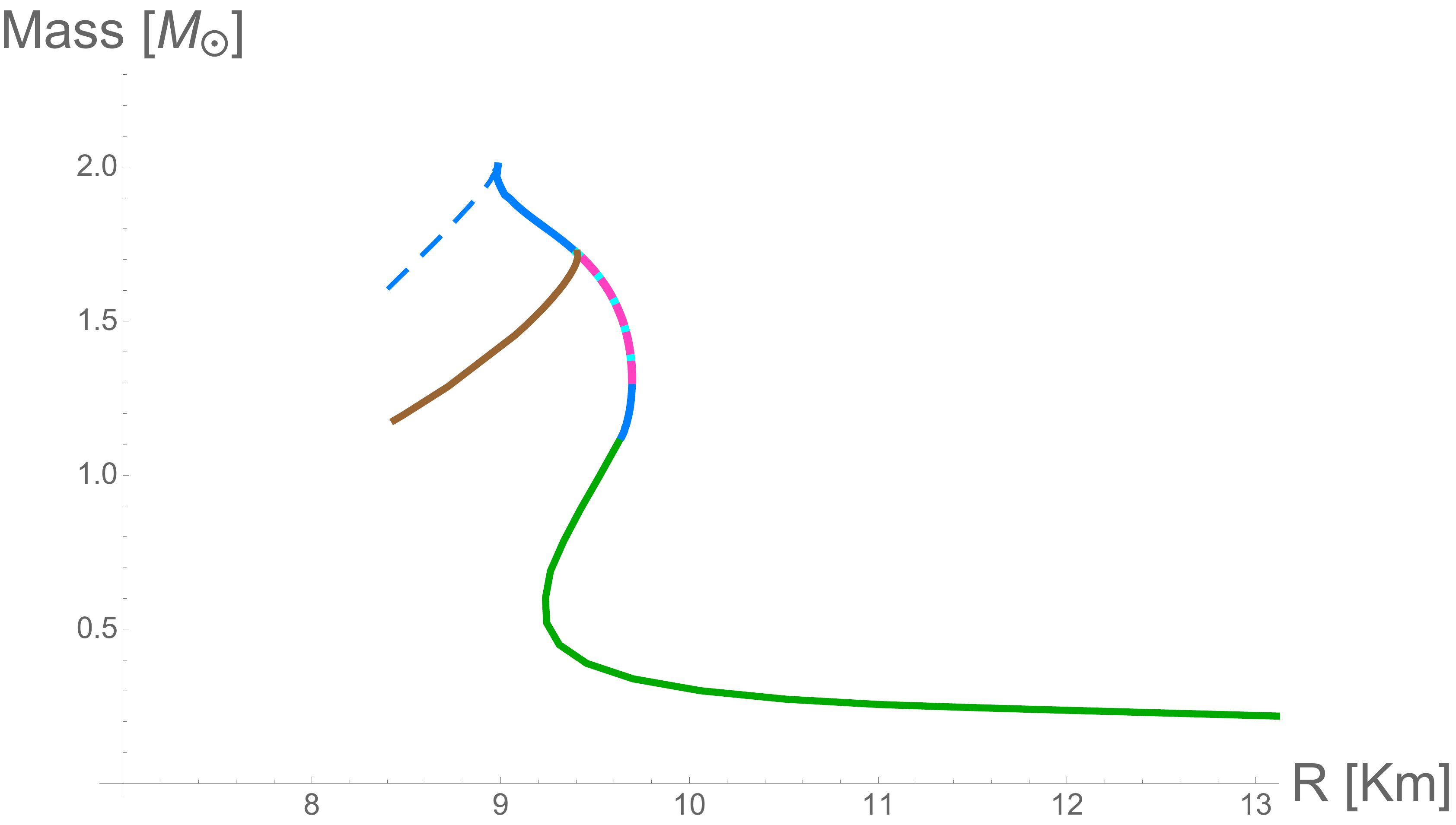}
    \includegraphics[width=9cm]{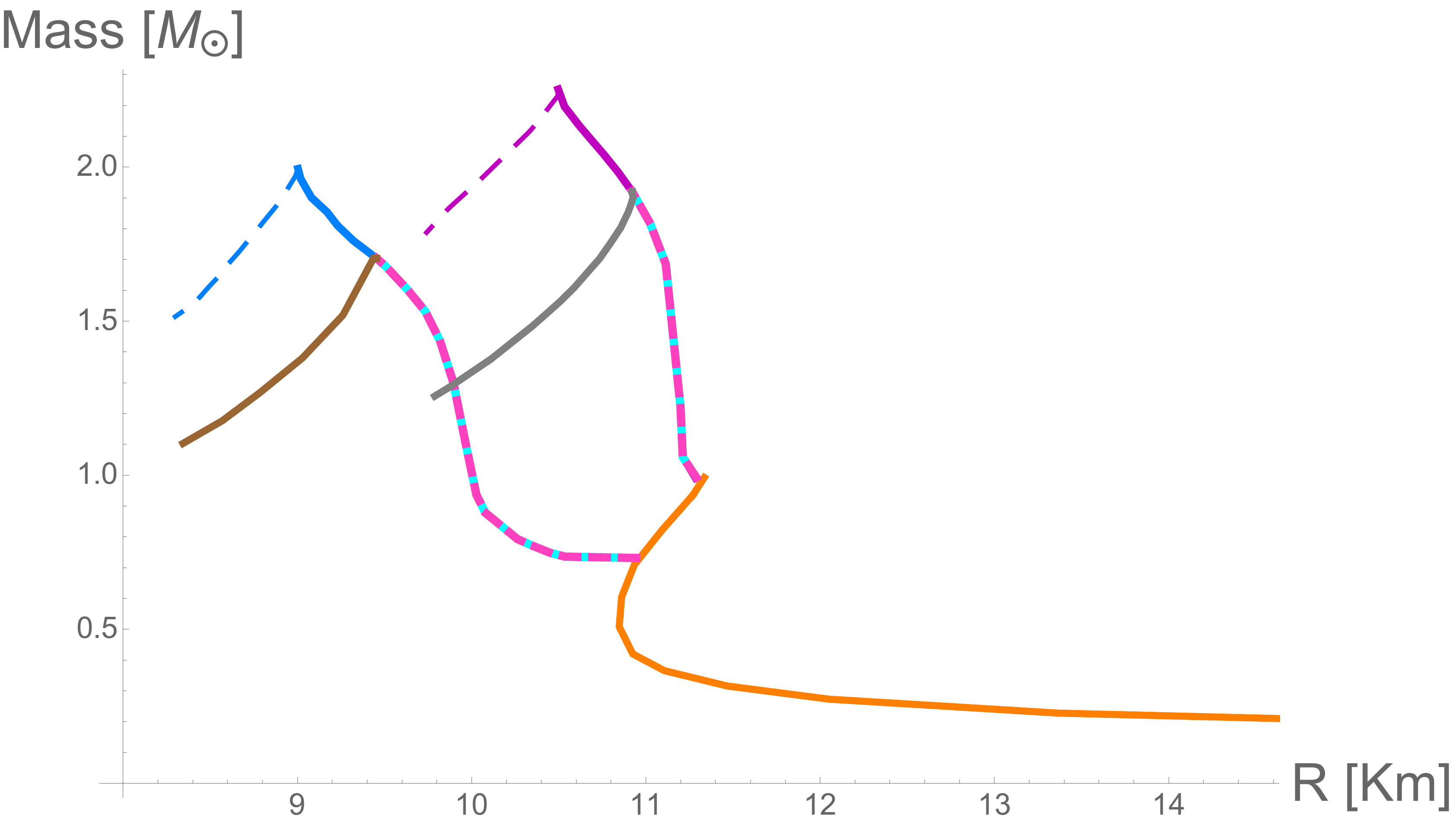}
    \includegraphics[width=9cm]{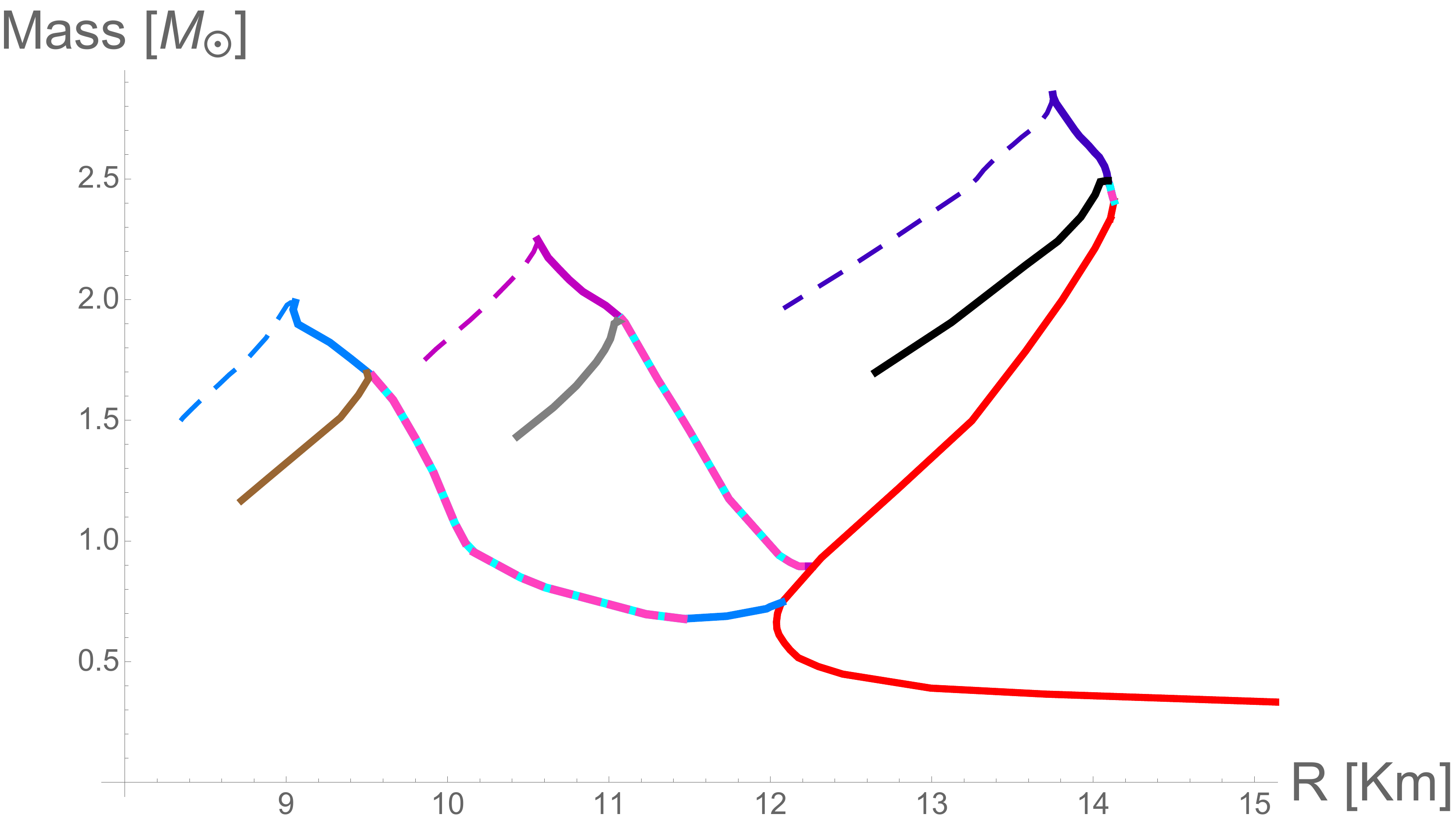}

 \noindent \textit{Figure 11. Mass vs radius curves for the case of $k_{IR}=0.575$ (blue),  $k_{IR}=0.35$ (magenta) and $k_{IR}=0.1$ (purple). The 
curves leaving the green/red/orange nuclear EoS prediction is the transition to a quark phase from Figure 10. The case with  $k_{IR}=0.35$ only has a transition from medium (orange) and stiff nuclear matter (red) and  the case with  $k_{IR}=0.1$ only has a transition from stiff nuclear matter. The stable branch where $c^2_s\leq1$ is indicated in dashed cyan/pink . The transition to a superconducting state for $\kappa=0.85$(brown), $\kappa=0.89$(gray) and $\kappa=0.94$(black) just before the \\speed of sound goes beyond 1 is also shown. }
\end{center}

\begin{center}
\includegraphics[width=8cm]{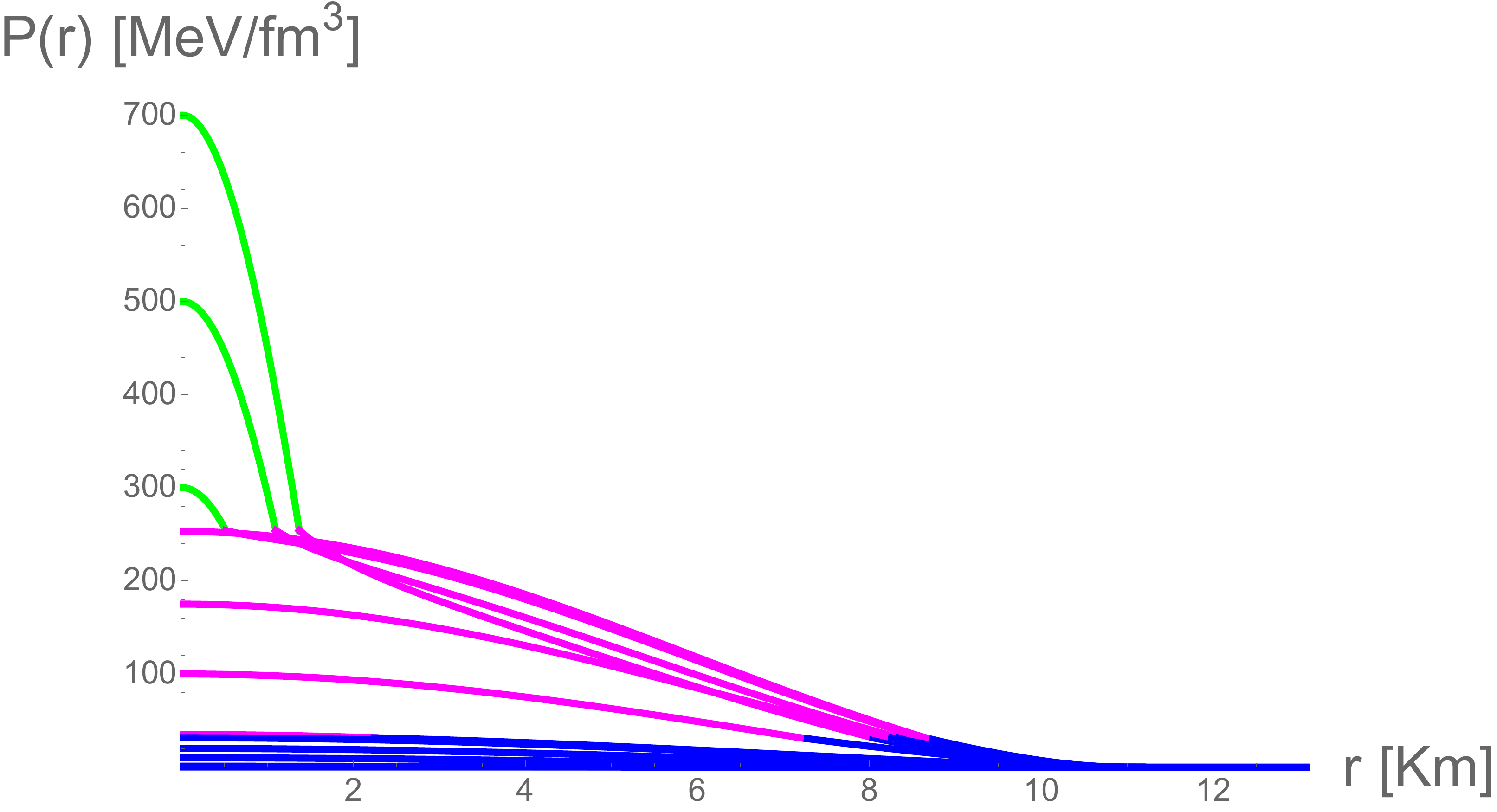}  
 
\noindent  \textit{ Figure 12. Pressure as a function of the radial distance $r$ in km for a hybrid star. The blue line corresponds to medium EoS nuclear matter. The
pink line corresponds to the massive chirally broken phase wih  $k_{IR}=0.35$. The
green line corresponds to the superconducting phase with $\kappa=0.89$ (stars with superconducting cores are unstable).}
\end{center}

Let's look at the top plot in detail as an example with the softest nuclear EoS. Here we only had the case $k_{IR}=0.575$ where the deconfined massive quark phase became the vacuum. To invoke a transition to the superconducting phase when the $c_s^2$ has just risen to one we set $\kappa=0.85$ - see the top two plots of Figure 10. The top plot of Figure 11 shows the resulting stars. Where the curve is green the nuclear phase only plays a role - the  star is neutrons to the core. The blue line marks where the  star has
    begun to have a deconfined massive quark phase in its core. If we did not include the superconducting phase but instead as in Figure 6 transitioned to the chirally restored $\chi=0$ phase this branch extends to the highest  point. After the core of the star experiences the transition to the chirally restored phase the stars become unstable - this is the sharp transition to the blue dashed line that angles down to the left. The region of these stars which satisfy the stability criteria above and have $c_s^2<1$ are 
marked by the dashed section of line. Finally if we allow a transition to the superconducting phase rather than the $\chi=\psi=0$ phase then we obtain the brown line - these stars with superconducting cores are again unstable but now the transition to them occurs at $c_s^2=1$ in the  deconfined massive quark phase, leaving a fully sensible picture of the dynamics at all $\mu$. This EoS does not support neutron stars as high in mass as the observational $\sim$2 solar mass limit so is presumably not a good description of QCD. 

In the central figure of Figure 11 we show example cases using the medium stiffness EoS for the nuclear phase. Here there are  deconfined massive quark phases for lower $k_{IR}$ and we show the cases of 0.575 and 0.35.  The plots show the same structure and elements as for the top plot as we have described. Again there are stars with quark cores but still reaching two solar masses is a struggle.

Finally in the bottom picture we show three cases for the stiffest nuclear EoS - $k_{IR}=0.575,0.35$ and 0.1. Between the last two of these values we find solutions with  deconfined massive quark cores and a upper most mass for stable stars between 2 and 2.5 solar masses. This is a considerable success. We have taken sensible phenomenological holographic models of the QCD EoS and shown that such stars can exist within sensible choices of parameters. This leads credence to the idea that quark cores can exist in neutron stars and hopefully encourages study for signals of such cores in gravitational wave signals from neutron star collisions. It is interesting to look at the structure of stars in this range. In Figure 12 we plot the pressure vs radius profiles of the stars for the case of the stiffest nuclear equation of state and $k_{IR}=0.35$, colouring the radial regions that are nuclear matter, deconfined massive quarks and the colour superconducting phase (stars with superconducting cores are unstable). We see that the quark core of some stars can be quite substantial. 

\section{IV LIGO Constraints for Tidal Deformabilites}

It is expected that in a colliding binary system of two neutron stars, the tidal forces between the two objects would have a measurable effect in the gravitational wave signal that could be observed using gravitational wave detectors. In \cite{TheLIGOScientific:2017qsa} the measurement of this effect was reported as a limit given for the tidal deformabilities of the two stars involved in the merger. To calculate the tidal deformability for specific solutions of the TOV equations that represent a neutron star, we follow references \cite{Hinderer:2009ca,Postnikov:2010yn,Zhao:2018nyf}.

The tidal deformation between neutron stars in a binary system connects the EoS, that describe the matter inside neutron stars,  to the gravitational wave emission during the inspiral. It has been shown that a small tidal signature arises in the inspiral below $400$ Hz \cite{Flanagan:2007ix}. This signature amounts to a phase correction which can be described in terms of a single EoS dependent tidal deformability parameter $\bar{\lambda}^{(\mathrm{tid})}$, which is the ratio of each star's induced quadrupole moment to the tidal field of its companion in the binary system. The parameter $\bar{\lambda}^{(\mathrm{tid})}$ depends on the EoS via both the neutron star radius $R$ and mass $M$, and a dimensionless quantity $k_{2}^{(\mathrm{tid})}$ called the Love number: 
\begin{equation}\label{tidaldefo}
\bar{\lambda}^{(\mathrm{tid})}= \frac{2}{3}\left(\frac{M}{R} \right)^{-5}k_{2}^{(\mathrm{tid})}
\end{equation}
A quick summary of this formalism is: one considers a static, spherically symmetric star of mass M placed in a time-independent external quadrupolar tidal field $\mathcal{E}_{i j}$.  In response, the star will develop a quadrupole moment $Q_{i j}$. In the star's local  rest frame, for large values of the radial coordinate $r$, the metric coefficient $g_{t t}$ is given by \cite{Thorne:1997kt}: 
\begin{equation}
\frac{\left(1-g_{t t}\right)}{2}=-\frac{m}{r}-\frac{3 Q_{i j}}{2 r^{3}}\left(n^{i} n^{j}-\frac{\delta^{i j}}{3}\right)+\frac{\mathcal{E}_{i j}}{2} x^{i} x^{j}+\ldots
\end{equation}
where $n^{i}=x^{i} / r$. This expansion defines the traceless tensors $\mathcal{E}_{i j}$ and $Q_{i j} .$ To linear order, the induced quadrupole will be of the form
\begin{equation}\label{quadrupole}
Q_{i j}=-\bar{\lambda}^{(\mathrm{tid})} \mathcal{E}_{i j}
\end{equation}
Thus $Q_{ij}$ and $\mathcal{E}_{ij}$ are defined as the coefficients in an asymptotic expansion of the metric at large distances from the star. 

The perturbation to the metric can be expanded in spherical harmonics. If one allows just the $l=2$ harmonic as a perturbation with a fixed spin axis then $\mathcal{E}$ and $\mathcal{Q}$ can be written in terms of $Y_{20}$. In the metric $Y_{20}$ is then multiplied by a function of $r$, $H(r)$. Now one solves the Einstein equations. A first degree differential equation is obtained for the radial function of the spherical harmonics $H(r)$ and, when solved together with the TOV equations, can give the value of the tidal deformability $\bar{\lambda}^{(\mathrm{tid})}$ from the following expression for the Love number $\kappa_2^{(\mathrm{tid})}$:
\begin{eqnarray}\label{lovenumber}
\kappa_{2}^{(\mathrm{tid})} &=&\frac{8}{5} \beta^{5}(1-2 \beta)^{2}[2+2 \beta(z_R-1)-z_R] \nonumber\\ &&
\left. \right. \hspace{0.3cm}\times\{2 \beta[6-3 z_R+3 \beta(5 z_R-8)] \nonumber \\ &&
+4 \beta^{3}\left[13-11 z_R+\beta(3 z_R-2)+2 \beta^{2}(1+z_R)\right] \nonumber \\ &&
\left.+3(1-2 \beta)^{2}[2-z_R+2 \beta(z_R-1)] \ln (1-2 \beta)\right\}^{-1}\nonumber \\ && 
\end{eqnarray}
where $\beta=M/R$ is the compactness parameter, and $z_R=RH'(R)/H(R)$ is evaluated using the surface value of the radial function $H(r)$ determined by the system of differential equations (\ref{radial-harmonic}):
\begin{eqnarray} \label{radial-harmonic}
\frac{d H}{d r}&=& Y \qquad \qquad \qquad\qquad \qquad \qquad\qquad \qquad  \qquad \quad \nonumber\\ 
\frac{d Y}{d r} &=& 2\left(1-2 \frac{M}{r}\right)^{-1} H\Big\{ -2\pi[5 \mathcal{E}+9 P+\frac{(\mathcal{E}+P)}{c^2_s}] \nonumber \\
&&+\frac{3}{r^{2}}+2\left(1-2 \frac{M}{r}\right)^{-1}\left(\frac{M}{r^{2}}+4 \pi r P\right)^{2} \Big\} \nonumber\\
&&+\frac{2 Y}{r}\left(1-2 \frac{M}{r}\right)^{-1}\left[-1+\frac{M}{r}+2 \pi r^{2}(\mathcal{E}-P)\right] \nonumber \\ && 
\end{eqnarray}
with boundary conditions $H(r)=a_0r^2$, $Y(r)=2a_0r$ for $r\ll1$.  $a_0$ is a constant that determines how much the star is deformed and can be fixed arbitrarily as it cancels in the expression for the Love number (\ref{lovenumber}). The functions $M(r)$, $P(r)$ and $\mathcal{E}(r)$ in  equation (\ref{radial-harmonic}) are the enclosed mass, pressure and energy density at the radial distance $r$, respectively, obtained using the TOV equations. Here we also have the appearance of the sound speed squared $c_s^2=\partial P/\partial \mathcal{E}$.
Thus by setting  the central pressure $P_c$, which fixes  the mass and the radius of the star, one can find the value of $z_R$. Then for each  $P_c$ we can obtain the  dimensionless tidal defomability of the star $\bar{\lambda}^{(\mathrm{tid})}$ as a function of the mass of the star.

LIGO and Virgo have provided a constraint for the value of the tidal deformability for a star of a  mass $M=1.4M_{\odot}$ \cite{TheLIGOScientific:2017qsa}, assuming slow rotation, at a 90$\%$ Bayesian probability level. In addition to this, Fig. 5 of \cite{TheLIGOScientific:2017qsa} gives both 90$\%$ and 50$\%$ probability contours for the independent tidal deformabilities of the two stars in the $\bar{\lambda}_1^{(\mathrm{tid})}-\bar{\lambda}_2^{(\mathrm{tid})}$ plane. 
Thus one can compare the results for a modelled star with these values, and show  how exotic phases relate to these contours. The curves are generated by independently determining the tidal deformabilities for each of the stars involved in the merger, obtaining the possible mass pairs using the chirp mass of the event, $M = 1.188M_{\odot}$.

We will choose to study two example sets of hybrid stars we have found found above. The first set corresponds to a transition from the soft nuclear matter to the deconfined massive quark matter with $k_{IR}=0.575$ - the top $M$ vs $R$ relation in Figure 11 (green to blue lines); the second case corresponds to a transition from the medium nuclear matter to the deconfined massive quark matter with $k_{IR}=0.35$ - right hand $M$ vs $R$ curve in the middle plot of Figure 11. (orange to magenta lines). We choose these cases because they predict hybrid stars for $M_\odot=1.4$ and through a good part of the possible range of masses cited for the detected signal from a binary neutron star inspiral in \cite{TheLIGOScientific:2017qsa}. For these cases we then compute the tidal deformability.
We plot the tidal deformability as a function of the mass for these two cases in Figure 13. Note the discontinuity in the green line simply reflects the movement from one part of the nuclear piecewise function to another. The spike in the blue curve near 1.5$M_\odot$ is due to an apparently accidental cancellation of terms driving the denominator of (\ref{lovenumber}) small.
We find that broadly as we increase the central pressure for both cases the tidal deformabilty decreases as the mass increases (this can be cross checked for example against the results in \cite{Annala:2017tqz} which use the same nuclear equations of state). LIGO and Virgo provide the constraint $\bar{\lambda}^{(\mathrm{tid})}\left(1.4 M_{\odot}\right) \leq 800$ for the likely case of slowly rotating stars (the low-spin prior) at a $90 \%$ Bayesian probability level. The models satisfy this constraint.

Additionally Fig. 5 of \cite{TheLIGOScientific:2017qsa} gives both $90 \%$ and $50 \%$ probability contours for the independent tidal deformabilities of the two stars on the $\bar{\lambda}_1^{(\mathrm{tid})}-\bar{\lambda}_2^{(\mathrm{tid})}$ plane, where $1$ and $2$ 

\begin{center}
\includegraphics[width=8cm]{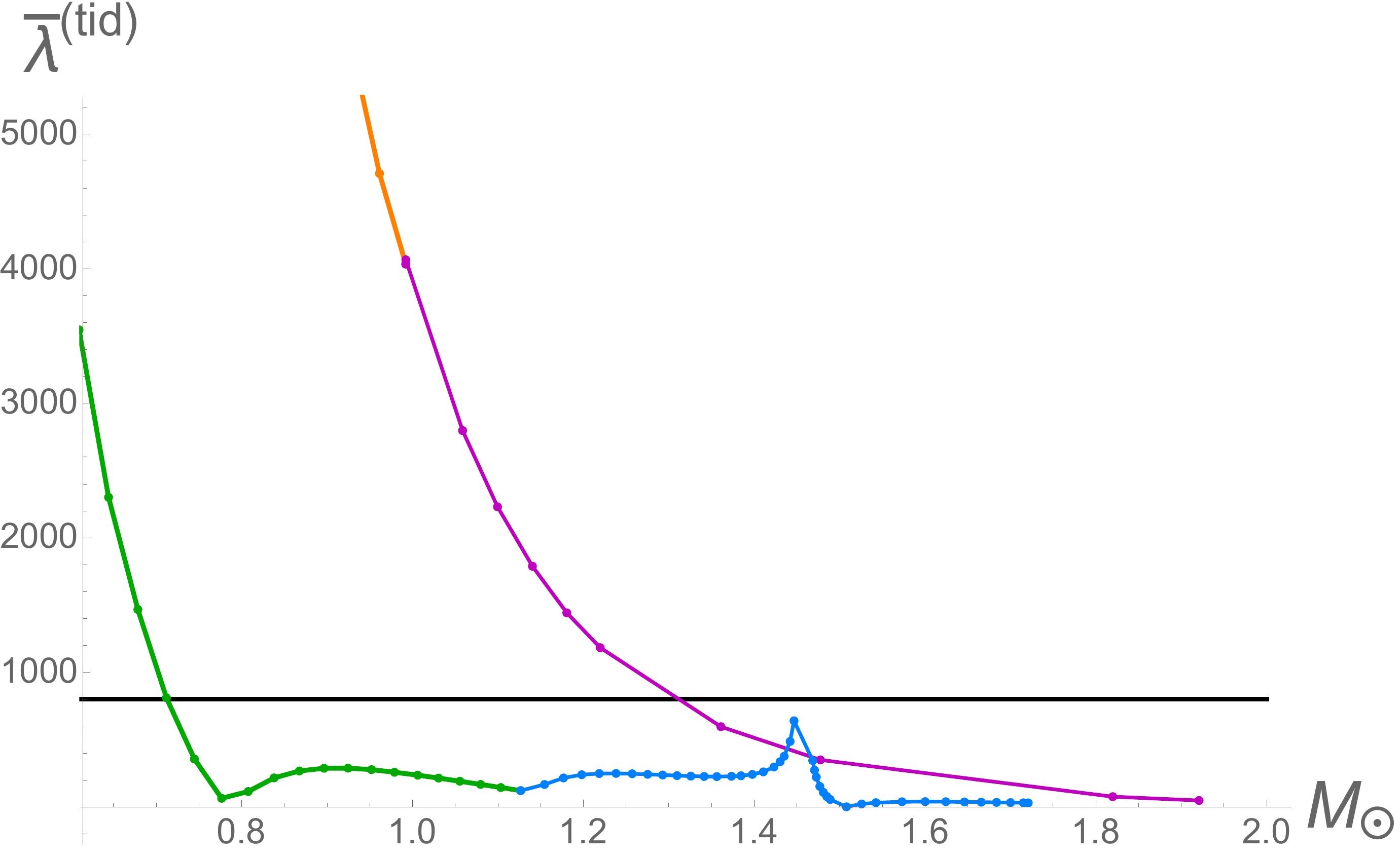}  
 
\noindent  \textit{Figure 13. The dimensionless tidal deformability as a function of the mass (in units of solar masses) for a holographic quark, and nuclear equations of state. The soft nuclear phase (green) has a transition to a chirally broken quark matter phase with $k_{IR}=0.575$ (blue).  The medium nuclear phase (orange) has a transition to a chirally broken quark matter phase with $k_{IR}=0.35$ (magenta). The LIGO/Virgo upper bound of $\bar{\lambda}^{(\mathrm{tid})}=$ 800 at 1.4$M_\odot$ is indicated by the horizontal line.}
\end{center}

\begin{center}
\includegraphics[width=8cm]{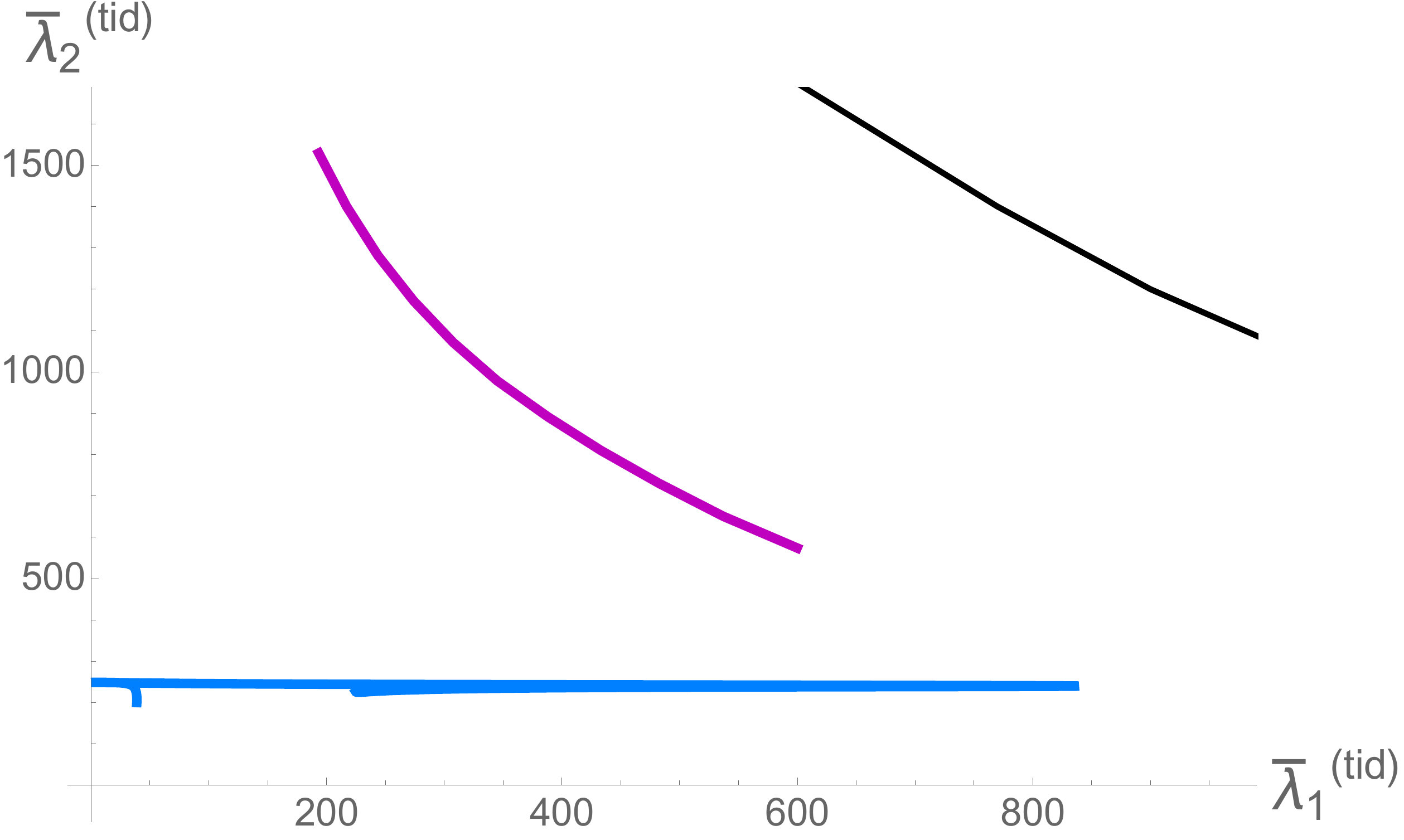}  
 
\noindent  \textit{Figure 14. The tidal deformabilities $\bar{\lambda}_{i}$ obtained for two stars with masses corresponding to those involved in the binary Neutron Star merger observed by LIGO and Virgo \cite{TheLIGOScientific:2017qsa}, corresponding to masses $m_{1} \in[1.36,1.60] M_{\odot}$ and $m_{2} \in[1.17,1.36] M_{\odot}$ (low-spin prior). The curves stand for the corresponding quark matter phases displayed in Fig. 13; the chirally broken quark matter phase has $k_{IR}=0.575$ (blue) and the chirally broken quark matter phase has $k_{IR}=0.35$ (magenta). The black curve is a sketch of the 90\% experimental bound contour given in figure 5 of \cite{TheLIGOScientific:2017qsa}.}
\end{center}

correspond to the two stars in the collision. To compare our results to these values, we show in Figure 14 how our example EoSs relate to these contours. The curves are generated by independently determining the tidal deformabilities for each of the stars involved in the merger.

To describe the binary system we take a chirp mass given by equation 
(\ref{chirpmass}), of $\mathcal{M}=1.188$ solar mass. 
\begin{equation}\label{chirpmass}
\mathcal{M}=\frac{\left(m_{1} m_{2}\right)^{3 / 5}}{\left(m_{1}+m_{2}\right)^{1 / 5}}
\end{equation}
where $m_{1}$ and $m_{2}$ are the masses of the components of the two body system. We can solve for $m_{1}$ in terms of $m_{2}$ for a fixed value of $\mathcal{M}$. Then we can vary $m_{2}$ as a parameter and use the relation we have for the tidal deformability as a function of  the mass of the star to obtain a relation between the tidal deformability for one of the two stars of the binary system with respect to the second one. This means that the the two stars involved in the binary neutron star merger correspond to masses $m_{1} \in[1.36,1.60] M_{\odot}$ and $m_{2} \in[1.17,1.36] M_{\odot}$. We observe that the two cases considered in Figure 15 fit inside the $90 \%$ probability contour given in \cite{TheLIGOScientific:2017qsa}. It is reasonable to hope that as more events are recorded that the bounds will begin to probe our models.

\section{V Discussion}

In this paper we have presented a holographic model of quark dynamics that provides a fully self consistent equation of state for quark matter and that allows for the existence of hybrid stars. The model has a sequence of transitions with $\mu$: the chirally broken vacuum first order transitions to a nuclear density phase above 308 MeV (we have not described this transition holographically); there is then a first order transition to a phase with a density of chirally broken  deconfined massive quarks; finally there is a first order transition to a chiral restored but colour superconducting phase. The speed of sound in each of the nuclear and deconfined massive quark phases grows with $\mu$ until the transition to the next phase (see Figure 10 for examples). Thus phases that resist second order transitions generate stiff matter. 

The holographic model of the quark dynamics has two regimes above and below the IR constituent quark mass. Above that scale the model is an AdS scalar dual to the chiral condensate with a radially dependent mass set by the running of the anomalous  dimension $\gamma$ in the gauge theory. When this running violates the BF bound in AdS a chiral condensate forms. At scales below the IR constituent quark mass we have needed to introduce a distinct description in the regime where the quarks should be integrated out - we have simply turned off the running mass in this very low energy regime which seems to provide a sensible IR completion of the model. The, very natural, discontinuity is though what drives the deconfinement transition to be first order (rather than the second order transitions we saw in smoother descriptions in \cite{Fadafa:2019euu}). 

In the chirally restored phase we have also introduced an AdS scalar to describe a colour superconducting Cooper pair condensate. Solutions with this present somewhat raise the pressure of the chirally restored vacuum.

Our model has three parameters: $\chi_0$ the IR quark mass that we have set to 330 MeV and sets the scale of the theory (this scale is formally introduced through the running coupling); $k_{IR}$ a parameter that weights the relative contributions to the action of the two regimes above and below the constituent quark mass; and $\kappa$ that controls the size of the interaction that triggers the superconducting condensate. The description of the nuclear phase taken from \cite{Hebeler:2013nza} also contains a range of EoS adding essentially an extra parameter through that choice. We have used the freedom of these parameters to construct a consistent set of phase transitions.  The resulting equations of state include stiff nuclear and then deconfined massive quark phases. 

We have solved the TOV equations using these equations of state to compute the structure of neutron/hybrid stars. We find hybrid stars with  deconfined massive quark cores for stars in the 1-2.5 solar mass range depending on the precise choice of parameters. The superconducting chirally restored phase is not stiff enough to support stars for sensible values of the condensate.

We also compared our stable solutions with observations made by the LIGO and Virgo collaboration of the tidal deformabilities obtained from the detection of gravitational waves from a binary neutron star inspiral. We observed an agreement with the data reported by LIGO and Virgo as the two cases we considered fit inside the 90\% probability contour.

We conclude that sensible holographic models can provide support to the idea that quark matter can be present at the cores of the most massive neutron stars observed. Our introduction of a deconfined massive quark phase requires a separation between chiral symmetry breaking and confinement at high density but this is far from impossible. Overall then this is an exciting conclusion that hopefully motivates further astrophysical and gravitational wave analysis of neutron stars and their collisions.

\noindent {\bf Acknowledgements:}

NEs work was supported by the STFC consolidated grant ST/P000711/1, and JCR's by Mexico's National Council of Science and Technology (CONACyT) grant 439332.


\end{document}